\documentclass[3p,11 pt]{elsarticle}

\usepackage{geometry}
\usepackage{relsize}
\usepackage{setspace}
\usepackage{mathtools}
\usepackage{amsmath}
\usepackage{amsfonts}
\usepackage{amssymb}
\usepackage{hyperref}
\usepackage{graphicx}
\usepackage[inkscapelatex=false]{svg}
\usepackage{subcaption}
\usepackage{epstopdf}
\usepackage{times}
\usepackage{subcaption}
\usepackage{natbib}
\usepackage{xcolor}
\usepackage{bbm}
\usepackage{dsfont}
\usepackage{amsthm}
\usepackage{stmaryrd}
\theoremstyle{definition}

\usepackage[ruled,vlined]{algorithm2e}
\hypersetup{colorlinks=true,linkcolor=blue,linkbordercolor=white,citecolor=blue}

\newcommand*{\defeq}{\mathrel{\vcenter{\baselineskip0.5ex \lineskiplimit 0pt
			\hbox{\scriptsize.}\hbox{\scriptsize.}}}%
	=}

\newcommand{\norm}[1]{\left\lVert#1\right\rVert}
\journal{International Journal of Engineering Science}
\begin{document}
	\begin{frontmatter}
	
	\title{A novel large-strain kinematic framework for fiber-reinforced laminated composites and its application in the characterization of damage}

    \author[mainaddress]{Shivam}
	\author[mainaddress]{Sandipan Paul\corref{correspondingauthor}}
	\ead{sandipan.paul@ce.iitr.ac.in}
	
 	\address[mainaddress]{Department of Civil Engineering, Indian Institute of Technology Roorkee, Roorkee 247667, India}
 \cortext[correspondingauthor]{Corresponding author}

\begin{abstract}

In this paper, a novel kinematic framework for fiber-reinforced composite materials is presented. {For this purpose, we use the theory of multiple natural configurations~(Rajagopal and Srinivasa, 1998, {\em International Journal of Plasticity}, \textbf{14}, 969-995) in conjunction with the multi-continuum theory (Bedford and Stern, 1972, {\em Acta Mechanica}, \textbf{14} (2), 85-102) keeping the underlying physics consistent.} The proposed kinematic framework results in a three-term decomposition of the deformation gradient, i.e., $\mathbf{F}=\mathbf{F}^e\mathbf{F}^r_\alpha\mathbf{F}^d_\alpha$, where $\alpha$ represents either the matrix or the fiber. After discussing the kinematic framework in detail, we use this new kinematic framework to characterize the damage contents associated with four damage mechanisms. These damage mechanisms are matrix cracking, fiber breakage, interfacial slip and debonding, and delamination. While the first two are derived by measuring the incompatibility of the pertinent configuration occupied by individual constituents, the latter two involve a relative displacement between either the constituents or the lamin\ae. The geometric interpretation {{of the developed damage measures}} is also presented using tools from differential geometry. The derived damage contents can be used in developing an appropriate constitutive model for laminated composites undergoing damage. 
 
\end{abstract}
	
	
\end{frontmatter}
	
\section{Introduction}

Fiber-reinforced laminated composites often exhibit complex material behaviors due to their composition. Being a multi-phase material, their damage mechanisms are often unique and different from other structural materials. {The commonly observed failure mechanisms include matrix cracking, fiber breakage, debonding and interfacial slip, fiber micro-buckling, interfacial sliding, and delamination, to name a few. Some of these mechanisms such as matrix cracking, fiber breakage, fiber micro-buckling etc. are associated with a particular phase of the material, viz., either the matrix or the fiber, while others involve both these phases. Since the 1970s, various theoretical models~\cite{aveston1971,garrett1977,hashin1985} based on the classical laminate theory have been developed to study the damage initiation and progress in composites. With the advent of new techniques in the field of theoretical and computational mechanics, more sophisticated models have been proposed. These models include but are not limited to geometry and physics-based models~\cite{talreja1985,olsson2001}, different numerical techniques such as finite element analysis~\cite{van2001,shetty2017}, phase field models~\cite{biner2009,bui2021} and, peridynamics~\cite{hu2011,roy2017} etc. These models are primarily focused on understanding the macroscopic behavior of the damaged material and the evolution of suitable damage variables. Collectively, these models are termed as continuum damage mechanics (CDM)~\cite{talreja1985,maire1997,maimi2007} in which continuum or volume-averaged damage measures for different mechanisms, viz., matrix crack density~\cite{talreja1985continuum,talreja1991continuum}, interfacial slip~\cite{hsueh1990interfacial,talreja1991continuum}, delamination~\cite{zou2001mode} etc., enter into the constitutive model through internal state variables. Since mechanical damage is essentially a dissipative process that results from the changes in the underlying microstructure, its evolution and effects on the macroscopic behavior of composites need to be investigated in greater details. For this purpose, a new set of models known as micro-damage mechanics (MIDM)~\cite{asp1996,kashtalyan2007} have been developed. These theories develop constitutive relations that account for the microscopic response and thereafter homogenization techniques are used to model the macroscopic behavior of the material. Apart from these two types, synergistic models~\cite{singh2009,montesano2015} have also been proposed to efficiently combine these two to enhance their applicability. On the experimental front, a number of new technologies have been developed that directly help in assessing damage in composite materials. These techniques include digital image and volume correlation techniques on different experimental setups such as fragmentation tests~\cite{garcea2017mapping}, fiber pull-out~\cite{tabiai2018situ}, end-notch fracture and mixed mode bending tests~\cite{zhu2020digital}, acoustic emission techniques~\cite{hamstad1986,saeedifar2020}, X-ray tomography~\cite{janicke2022debonding,drouhet20233d} etc. These methods are not only effective in assessing the current state of damage, but are also capable of capturing damage progression for both small and large deformations.}

Depending on their specific constituents, the damage mechanisms of laminated composites may widely vary. For example, composite materials made up of glass or carbon fibers and a resin-based matrix, which are widely used in civil engineering or aerospace applications, are stiff in nature. These composites show brittle failure and cannot sustain large deformation. On the other hand, fiber-reinforced composites that constitute soft matrices such as elastomers or rubbery materials can undergo large deformations without failure due to the specific mechanical properties of their constituents as well as their damage mechanisms. These materials find extensive applications in soft robotics~\cite{zhang2021soft}, deployable origami structures~\cite{deleo2020origami}, and biomedical applications~\cite{gasser2006hyperelastic}. {In spite of the large volume of literature on damage in composite materials, most of these models are limited to a small-strain theory as required for the former type. Although some damage models~\cite{tham2005,mandel2015,costa2022} have been proposed for finite deformations, models based on the intrinsic geometry of the material, configurational changes and associated thermodynamics is relatively scarce.} While some damage mechanisms, such as matrix cracking, fiber breakage, debonding and interfacial slip, delamination etc., are common to both types of composites, some specific damage mechanisms such as micro-buckling of fibers, are also important in the latter type~\cite{merodio2003instabilities,jimenez2014modeling}. In view of the widely varying material behaviors of fiber-reinforced composites, it is important to develop a suitable, sufficiently general continuum framework that can accommodate the necessary kinematic features allowing for large-strain damage. For this purpose, we use the idea of a multicontinuum theory, developed by Bedford and Stern~(1972)~\cite{bedford1972multi} in conjunction with the theory of multiple natural configurations, developed by Rajagopal and Srinivasa~(1998)~\cite{rajagopal1998mechanics,rajagopal2004thermomechanics} to propose a novel large-strain kinematic framework and utilize this to characterize different damage mechanisms in fiber-reinforced composites.

To accommodate large deformation in fiber-reinforced composites, many constitutive models have been proposed in the literature~\cite{merodio2003instabilities,nguyen2007modeling}. These models often consider the constituent phases of the composites together and develop constitutive models by using invariants such as $I_4$, $I_5$ etc. that take into account the fiber-orientations. A notable exception is the work of Bedford and Stern~(1972)~\cite{bedford1972multi} in which the constituents are modeled as individual continua and the overall response of the composite is obtained by considering it as a superimposition of these individual continua. This framework is referred to as a multicontinuum theory and closely resembles the idea of a mixture theory of fluids. Akin to the mixture theory, this work also models the interaction between the different phases, i.e., the matrix and the fibers, based on the relative displacement between them and their corresponding interaction forces. This theory has been further developed by Hansen and coworkers in the context of its finite element (FE) implementation~\cite {hansen1995multicontinuum}, failure analysis of laminates~\cite{mayes2004composite,nelson2012failure}, failure in woven composite~\cite{key2007progressive}, delamination~\cite{nelson2011delamination} etc. Although this framework was primarily developed within a large deformation setting, the subsequent works by Hansen and co-workers are limited to a linearized theory.

Another significant development in modeling dissipative processes within a finite deformation framework is the theory of multiple natural configurations, developed by Rajagopal and Srinivasa~(1998)~\cite{rajagopal1998mechanics}. Although the fundamental tenet of this theory is similar to a multiplicative decomposition of the deformation gradient, i.e., $\mathbf{F}=\mathbf{F}^e\,\mathbf{F}^i$, its interpretation is altered in the following way. When an infinitesimal neighborhood of a material particle is subjected to an instantaneous elastic unloading from the current configuration of the body, (i.e., all the external stimuli are removed) it occupies a local stress-free configuration, called the natural configuration. This natural configuration is related to the undeformed configuration through the tangent map $\mathbf{F}^i$. Physically, the tangent map $\mathbf{F}^i$ accounts for the microstructural changes under a dissipative process. Since the underlying microstructure of the body evolves during a dissipative process, an elastic unloading at different time instants will lead to different natural configurations. Thus, a body is considered to possess multiple natural configurations when subjected to a dissipative process. While the overall response of the body can be viewed as a family of elastic responses measured from a given natural configuration, the evolution equation for these natural configurations is obtained by employing appropriate thermodynamic restrictions, in this case, a maximum rate of dissipation criterion~\cite{rajagopal1999thermomechanics}. This framework has been successfully employed in a variety of problems, such as viscoelasticity and rate-independent plasticity~\cite{rajagopal1998mechanics,paul2021constitutive,singh2025extension}, shape memory materials~\cite{rajagopal1999thermomechanics}, mechanics of polymers~\cite{song2019thermodynamically,wijaya2025mixture}, to name a few. 

Motivated by these major works, we extend this framework to model fiber-reinforced laminated composites undergoing damage. Here, we use the following modeling strategy. We start with the motion of a composite particle consisting of both fibers and matrix. Following the theory of multiple natural configurations, when an infinitesimal neighborhood around the composite particle is subjected to an instantaneous elastic unloading through $\mathbf{F}^{e^{-1}}$, it occupies a local natural configuration. Now, in view of the multi-continuum theory of Bedford and Stern~(1972)~\cite{bedford1972multi}, the composite particle containing both the matrix and fiber is considered as a single particle due to the interaction force between the constituents. To understand the underlying mechanics, however, the composite particle in the natural configuration is further subjected to another unloading in which these interaction forces are removed.  As a result, the inelastic deformation gradient, $\mathbf{F}^i$ is further decomposed into two parts, viz., $\mathbf{F}^i=\mathbf{F}^r_{\alpha}\,\mathbf{F}^d_{\alpha}$ where $\alpha$ represents either the matrix or the fiber. Physically, this decomposition is similar to the unloading process of Rajagopal and Srinivasa~(1998)~\cite{rajagopal1998mechanics} in which the tangent map $\mathbf{F}^{r^{-1}}_{\alpha}$ results in two separate natural configurations for the composite constituents and these configurations are related to the undeformed configuration through the tangent maps $\mathbf{F}^d_{\alpha}$. Thus, the total deformation gradient is multiplicatively decomposed into three terms, viz., $\mathbf{F}=\mathbf{F}^e\,\mathbf{F}^r_{\alpha}\,\mathbf{F}^d_{\alpha}$.{A three-term multiplicative decomposition of the deformation gradient has been previously used in different contexts such as, polycrystalline plasticity~\cite{kratochvil1972finite,bammann2001model}, viscoelasticity~\cite{malek2018derivation} as well as thermo-viscoelasticity of multi-phase polymer mixtures~\cite{sreejith2023thermodynamic}. However, the underlying physics of the proposed framework is particularly developed for characterizing the damage mechanisms in fiber-reinforced laminated composites from a theoretical mechanics perspective.}

{The primary objective of this paper is to develop a kinematic framework within the theory of configurational mechanics based on the aforementioned modeling strategy. Unlike the traditional finite-deformation models for composites, the developed framework takes into account the motions of the individual constituents as well as the interactions between them. Therefore, this framework not only facilitates the geometric characterization of different damage mechanisms, but also enables the development of constitutive models that account for their evolution and effects on the overall response of the material, thereby providing new insights into the underlying physics. While the latter is an important future direction of research, the current paper focuses on developing the kinematic framework as well the characterization of different damage mechanisms from physical as well geometric perspectives.} In particular, we study the geometric features of the relevant configurations to characterize the damage contents in terms of the relevant tangent maps. We consider four damage mechanisms: matrix cracking, fiber breakage, debonding and interfacial slip, and delamination. A multiplicative decomposition of the deformation gradient has been useful in characterizing various material defects, especially in the context of plasticity. Several incompatibility measures of the relevant configurations have been proposed to characterize geometrically necessary dislocations~\cite{acharya2001model,paul2020characterizing} as well as other defects such as disclinations, deformation twinning etc.~\cite{clayton2010nonlinear,clayton2014alternative}. In the context of damage mechanics, a similar method was used by Kachanov~(1980)~\cite{kachanov1980continuum} to obtain a suitable measure which he termed as the \emph{crack density tensor}. In this paper, we implement these ideas to our framework to characterize matrix cracking and fiber breakage. For an incoherent phase transition problem, Cermelli and Gurtin~(1994)~\cite{cermelli1994kinematics} proposed a method to measure the relative slip or the tangential mismatch between different phases. This method is found to be useful in our work to characterize debonding and interfacial slip. For the purpose of characterizing delamination, we develop a method similar to the work of Gupta and Steigmann~(2012)~\cite{gupta2007evolution,gupta2012plastic} where they considered the problem of a solid with an interface undergoing plastic deformation. They introduced a new measure of material defects in terms of the interface dislocation density. Inspired by these developments, we utilize the idea of relative slip, interface, and the associated incompatibilities to model delamination in the fiber-reinforced composites.  

The rest of the paper is organized as follows. In \S~\ref{sec:Preliminaries}, the relevant kinematics and mathematical preliminaries are discussed. The kinematic framework is developed for composite materials in \S~\ref{sec:Three-term decomposition in composites}, where the central idea from the multi-continuum theory and multiple natural configurations are incorporated. Using this framework, we present the characterization of different damage mechanisms, particularly matrix cracking, fiber breakage, interfacial debonding and fiber pull-out, and delamination in \S~\ref{Damage characterization}. The geometric interpretation of these damage measures is provided in \S~\ref{sec:Geometric interpretation of the measures of damage}. {\S~\ref{sec:discussion} is devoted to discussions on three topics related to the developed damage measures-- a general constitutive framework in which these measures can be incorporated, their relations with the existing damage variables and, possible experimental methods for their quantification.} Finally, the findings of this paper are summarized and the paper is drawn to a conclusion. Throughout the paper, letters with $~\tilde{(\cdot)}~$ are used for the interface. ${D(\cdot)}/{Dt}$ denotes a material time derivative. {For a continuous field,} `Curl' of a vector field $\mathbf{v}$  with respect to the Cartesian, reference (or material) coordinates is defined as $(\text{Curl}\,\mathbf{v})_i=\epsilon_{ijk}\,{\partial{v}_k}/{\partial{X}_j}$, while with respect to the current (or spatial) coordinates, we define $(\text{curl}\,\mathbf{v})_i=\epsilon_{ijk}\,{\partial{v}_k}/{\partial{x}_j}$. {For a discontinuous vector field $\mathbf{v}$, a jump in the vector field across an interface is defined as $\llbracket\mathbf{v}\rrbracket\defeq \mathbf{v}^+ -\mathbf{v}^-$. The gradient of this vector with respect to the reference coordinates is defined as $(\nabla_{\mathbf{X}}\mathbf{v})_{ij}\defeq{\partial{v}_i}/{\partial{X}_j}=\left({\partial{v}_i}/{\partial{X}_j}\right)_{regular} +(\llbracket {v}\rrbracket_i\, {N}_j\,\delta)_{jump}$ where $\mathbf{N}$ is the normal to the interface and $\delta$ denotes the Dirac delta measure. In a similar manner, `Curl' of the vector field with respect to the reference (or material) coordinates is defined as $(\text{Curl}\,\mathbf{v})_i=(\epsilon_{ijk}\,{\partial{v}_k}/{\partial{X}_j})_{regular}+(\epsilon_{ijk}\,\llbracket {v}_k\rrbracket {N}_j\,\delta)_{jump}$.} Partial derivatives with respect to spatial coordinates is written in shorthand as $\partial_B(\cdot)={\partial(\cdot)}/{\partial X^B}$.

\section{Preliminaries}\label{sec:Preliminaries}

In this section, we briefly revisit the idea of multiple natural configurations and other necessary topics that will be used in the characterization of material defects.
\subsection{Kinematics}\label{sec:Kinematics}
Let us consider a body $\mathbf{\mathcal{B}}$ embedded in a Euclidean point space. At time $t=0$, the body occupies an undeformed configuration $\kappa_R(\mathcal{B})$ and the position vector of a material particle of the body in this configuration is denoted by $\mathbf{X}$. A motion $\mathbf{x}=\mathcal{X}(\mathbf{X},t)$ maps each material particle of the body from its undeformed (reference) configuration to its current (deformed) configuration at time $t$, denoted by $\kappa_t(\mathcal{B})$. Here $\mathbf{x}$ represents the position vector of the corresponding material particle in the current configuration of the body. Let $d\mathbf{X}$ and $d\mathbf{x}$ denote the infinitesimal fibers in the undeformed and current configurations, respectively. A deformation gradient $\mathbf{F}$ is a tangent map that takes a tangent vector from the undeformed configuration of the body and places it into the tangent space of its current configuration. The deformation gradient, thus, can be written as 
\begin{equation}\label{eq:deformation_gradient}
     d\mathbf{x}=\mathbf{F}\,d\mathbf{X}\quad \text{with}\quad \mathbf{F}(\mathbf{X},t)\defeq \dfrac{\partial{\mathcal{X}}(\mathbf{X},t)}{\partial \mathbf{X}}. 
\end{equation}
The tangent map $\mathbf{F}$ can be written as a gradient of a {vector} function, as shown in the second part of Eq.~\eqref{eq:deformation_gradient}, since the undeformed and the current configurations of the body are globally compatible, i.e., $\text{Curl}(\mathbf{F})=\mathbf{0}$. Based on this deformation gradient, the right Cauchy Green tensor and the Green strain tensor are defined as
\begin{equation}
    \label{eq:right_cauchy}
    \mathbf{C}=\mathbf{F}^T\,\mathbf{F},\quad \mathbf{E}=\dfrac{1}{2}(\mathbf{C}-\mathbf{I}).
\end{equation}
The velocity and velocity gradient, associated with this motion, are defined as 
\begin{equation}\label{eq:velocity}
     \mathbf{v}({t})\defeq \frac{D {\mathcal{X}}(\mathbf{X},t)}{D{t}},\quad \mathbf{L}\defeq\nabla_x\mathbf{v}=\dot{\mathbf{{F}}}\,\mathbf{F}^{-1}.
\end{equation}
\begin{figure}[h]
      \centering
      \includegraphics[width=0.75\linewidth]{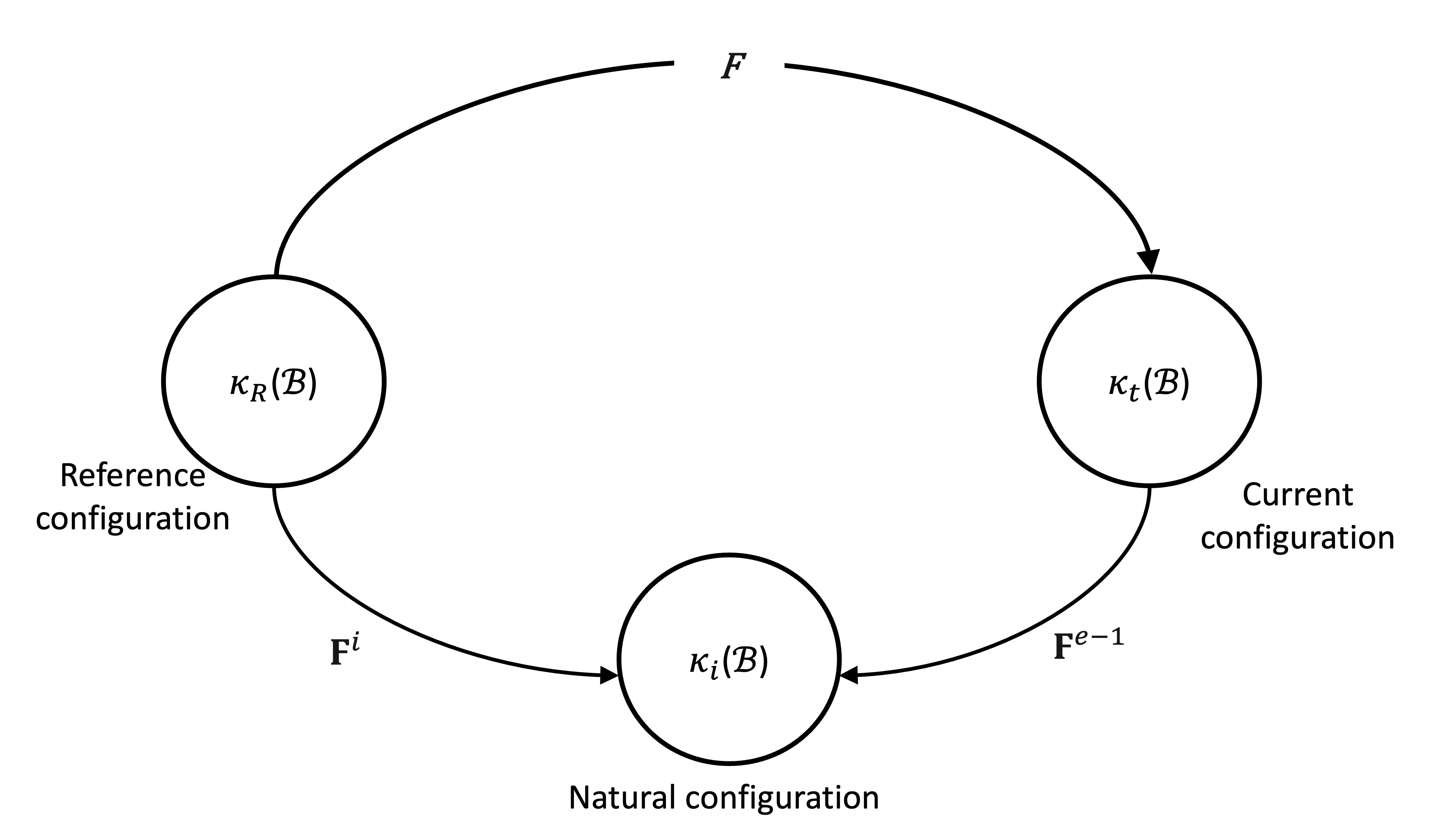}
      \caption{A multiplicative decomposition of the deformation gradient and the relevant configurations.}
      \label{fig:two-term decomposition}
  \end{figure}
  
To model an inelastic process, a multiplicative decomposition of the deformation gradient is typically used in the form of
\begin{equation}\label{eq:elastic_plastic_decomposition}
     \mathbf{F} =\mathbf{F}^e\,\mathbf{F}^i.
\end{equation}
The physical significance of this decomposition can be understood through the multiple natural configurations framework, developed by Rajagopal and Srinivasa~(1998)~\cite{rajagopal1998mechanics}. According to this framework, when an infinitesimal neighborhood around a material particle of the body is subjected to an instantaneous elastic unloading from its current configuration $\kappa_t({\mathcal{B}})$ through $\mathbf{F}^{e^{-1}}$, the body occupies a locally stress-free configuration $\kappa_{i}({\mathcal{B}})$, called the natural configuration. This natural configuration is related to the undeformed configuration of the body $\kappa_R({\mathcal{B}})$ through the inelastic part of the deformation gradient, $\mathbf{F}^i$. These configurations and the corresponding tangent maps are shown in Fig.~\ref{fig:two-term decomposition}. It is important to note that the natural configuration is obtained through a \emph{local} instantaneous unloading of the neighborhoods of a material particle of the body. Due to this local nature of the natural configuration, it is not, in general, globally compatible. The incompatibility of the natural configuration has been used to characterize different material defects, particularly in the context of plasticity. This incompatibility can be measured by~\cite{cermelli2001characterization} 
\begin{equation}
    \mathbf{b}^i=\int_{\partial\Omega}\mathbf{F}^id\mathbf{X}=\int_{\Omega}(\text{Curl}\,\mathbf{F}^i)^T\,{\mathbf{N}}\,\mathrm{d}{A}
    \label{eq:incompatibilityFiFe1}
\end{equation}
where ${\mathbf{N}}\,\mathrm{d}{A}$ represents the infinitesimal vector area within the region $\Omega\subset\kappa_R(\mathcal{B})$. The measure of incompatibility is further modified by pushing forward the vector area from the reference configuration of the body to its natural configuration. This transformation results in 
\begin{equation}\label{eq:Burgers_vector}
    \mathbf{b}^i=\int_{\Omega}\dfrac{1}{J^i}(\text{Curl}\,\mathbf{F}^i)^T\,\mathbf{F}^{i^T}\,\overline{\mathbf{n}}\,\mathrm{d}\overline{a}.
\end{equation}
Here $\overline{\mathbf{n}}\,\mathrm{d}\overline{a}$ is the infinitesimal vector area in the natural configuration of the body and the Jacobian, $J^i=\text{det}(\mathbf{F}^i)$.

\subsection{Discontinuity in a displacement field}\label{sec:Continuous and discontinuous manifold}

In \S~\ref{sec:Kinematics}, the body $\mathcal{B}$ is considered to be a smooth differentiable manifold. This continuity is disrupted when the underlying microstructure of the body is altered by the introduction of a microcrack during a dissipative process. From a geometric perspective, this discontinuity can be characterized by a non-vanishing local torsion of the material manifold. To understand this, we follow the framework proposed by Valanis and Panoskaltsis~(2005)~\cite{valanis2005material}. Let us take a closed circuit $\text{PQRS}$ in the undamaged body $\mathcal{B}$ as shown in Fig.~\ref{fig:Crack}. In the undamaged body, there are two possible paths $\text{RQP}$ and $\text{RSP}$ that one can take to reach point $\text{P}$ from point $\text{R}$. In this configuration, $\text{P}$ is a single-valued point which is independent of the path followed. But whenever a crack exists in the body, the field is not continuous anymore. Now if we follow the same paths again in the deformed configuration of the body, $\text{P}$ is no longer a single-valued point, and thus the circuit $\text{PQRS}$ is not closed. In this case, we arrive at two different points, $p_1$ and $p_2$. The jump in the circuit can be defined as 
\begin{equation}\label{eq:jump_in_p}
    \mathbf{p}=\mathbf{p}_1-\mathbf{p}_2
\end{equation}
where $\mathbf{p}_1$ and $\mathbf{p}_2$ are the position vectors of the respective points. {Valanis and Panoskaltsis~(2005)~\cite{valanis2005material} showed that (see~\ref{sec:Appendix_discontinious_field} for a detailed derivation) this local discontinuity can be written in terms of the deformation gradient $\mathbf{F}$ as 
\begin{equation}\label{eq:curl}
    \mathbf{p}=(\text{Curl}\,\mathbf{F})\,d\mathbf{A}
\end{equation}
where $d\mathbf{A}$ is the vector crack surface area.
\begin{figure}[h]
    \centering
    \includegraphics[width=0.6\linewidth]{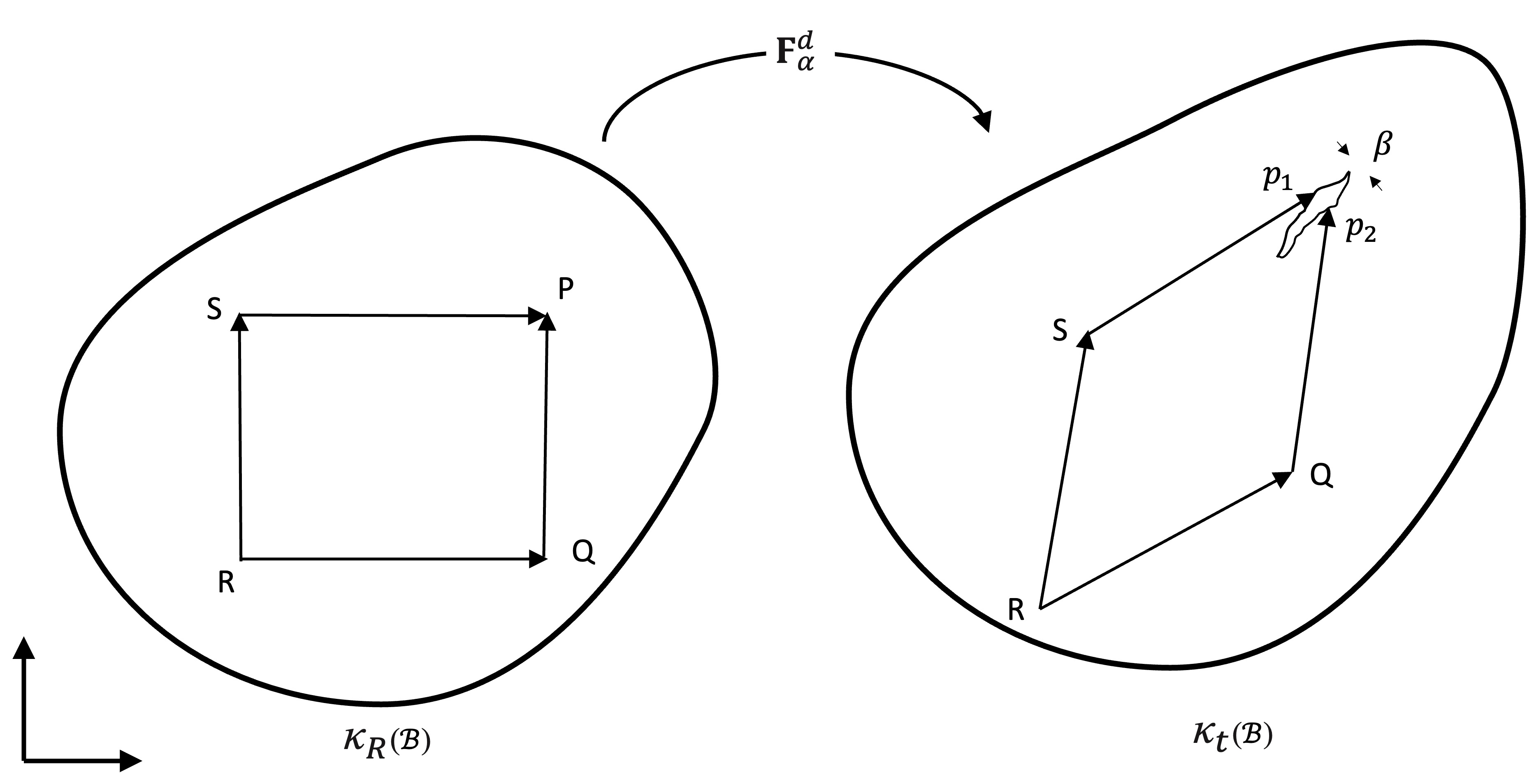}
    \caption{A motion corresponding to $\mathbf{F}$ maps the undeformed configuration of the body to its deformed configuration where a (micro-) crack exists. Due to discontinuity in the displacement field in the presence of a crack, P is no longer a single-valued point, and thus, the circuit PQRS is not closed in $\kappa_t(\mathcal{B})$.}
    \label{fig:Crack}
\end{figure}

Now following Kachanov's~(1980)~\cite{kachanov1980continuum} definition, a crack in a continuum may also be defined in terms of the jump in the displacement field, $\llbracket\mathbf{u}\rrbracket=\mathbf{u}^+ -\mathbf{u}^-\defeq\mathbf{q}$ and, the normal $\mathbf{n}$ to the crack surface $\text{S}$. For a crack with normal discontinuity, the vectors $\mathbf{q}$ and $\mathbf{n}$ are parallel to each other and hence, one can write $\mathbf{q}=d\,\mathbf{n}$, with $d$ being the distance between the bounding surfaces of the crack in the manifold as shown in Fig~\ref{fig:Crack}. Based on this framework, Kachanov~(1980)~\cite{kachanov1980continuum} defined the crack density tensor as
\begin{equation}\label{eq:crack_distortion_tensor}
    \boldsymbol{\alpha}\defeq \mathbf{q}\otimes\mathbf{n}\,\delta(\text{S})=d\,\mathbf{n}\otimes\mathbf{n}\,\delta(\text{S}).
\end{equation}
Here $\delta$ is the field concentration on the surface $\text{S}$. Note that this definition was also used by Talreja~(1985)~\cite{talreja1985continuum} in his characterization of matrix crack density in laminated composites. From the above discussion, it is evident that the displacement jump $\mathbf{q}$ is sufficient to characterize the local geometry of the defect as $\mathbf{q}=d\,\mathbf{n}$. Moreover, since $\mathbf{p}$ and $\mathbf{q}$ both represent the jump in the displacement field due to a crack, one can write
\begin{equation}\label{eq:Valanis_kachanov}
    \mathbf{q}=d\,\mathbf{n}=(\text{Curl}\,\mathbf{F})\,d\mathbf{A}.
\end{equation}
}
This definition of a discontinuity is capable of accommodating different kinds of material incompatibility irrespective of their specific physical nature. In this paper, we employ a notion similar to the one used by Valanis and Panoskaltsis~(2005)~\cite{valanis2005material} to characterize various damage mechanisms in fiber-reinforced laminated composite materials.  

\subsection{Interface}\label{sec:Interface}
\begin{figure}[h]
    \centering
    \includegraphics[width=0.7\linewidth]{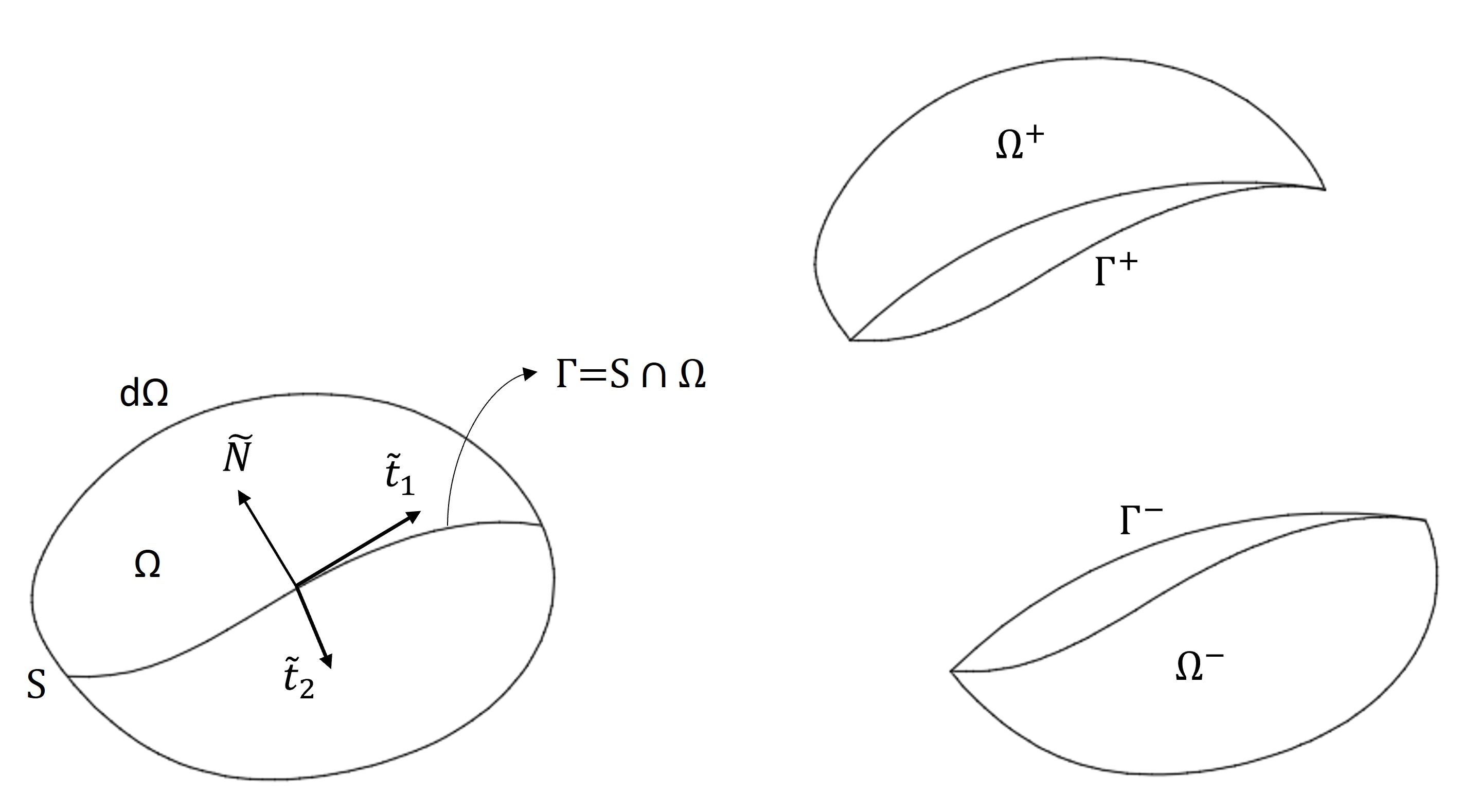}
    \caption{Interface $S$ in a region of the continuum $\Omega$. $\Gamma$ represents a curve that represents the intersection between the interface and the bulk of the continuum, i.e., $\Gamma=S~\cap~\Omega$. The body can be divided into two parts across the interface, denoted by the signs `+' and `-'. $\tilde{t}_1$ and $\tilde{t}_2$ are mutually orthogonal base vectors on the 2-D interface whereas $\tilde{N}$ is normal to these base vectors. }
    \label{fig:Interface}
\end{figure}

In this section, we revisit the characterization of defects across an interface that will later be used for measuring delamination in laminated composites. The development here closely follows the formulation of Gupta and Steigmann~(2012)~\cite{gupta2012plastic}, and more details can be found there. The interface (or a singular surface) in a continuum is a region across which a jump can be observed in a field (such as $\mathbf{\Phi}$) that is continuous across the rest of the continuum. An interface can be viewed as a two-dimensional manifold embedded in the reference configuration of the body $\kappa_R({\mathcal{B}})$. Let $\Omega$ be a region within the reference configuration of the body, bounded by the surface $\partial\Omega$. Let $S$ denote the interface and $\Gamma$ be a curve that represents the intersection between the interface $S$ and the rest (bulk) of the continuum, $\Omega$, as shown in Fig.~\ref{fig:Interface}. Let $\tilde{\mathbf{{t}}}_1$ and $\tilde{\mathbf{t}}_2$ be a set of orthogonal bases in the tangent space of the interface and let $\tilde{\mathbf{N}}$ denote the normal to the interface such that $\tilde{\mathbf{N}}=\tilde{\mathbf{t}}_1\times \tilde{\mathbf{t}}_2$. As shown in Fig.~\ref{fig:Interface}, the bulk of the continuum can now be divided into two parts across the interface: $\Omega^+$ and $\Omega^-$. Therefore, one can further define $\Gamma^{+/-}=S\,\cap\,\Omega^{+/-}$ such that $\Gamma=\Gamma^+\,\cup\,\Gamma^-$. 
The jump in a field $\mathbf{\Phi}$ can be defined as 
\begin{equation}\label{eq:jump}
    \llbracket\mathbf{\Phi}\rrbracket= \mathbf{\Phi}\,^{^+} -\mathbf{\Phi}\,^{^-},
\end{equation}
The total incompatibility in the natural configuration of the body can be written as~\cite{gupta2007evolution}
\begin{equation} \label{eq:incompatibilityFi}
\int_{\partial\Omega}\mathbf{F}^i d\mathbf{X}=\int_{\Omega}(\text{Curl}\,\mathbf{F}^i)^T\,\mathbf{N}\,dA-\int_\Gamma\llbracket\mathbf{F}^i\rrbracket d\mathbf{X}.
\end{equation}
One can observe that the first term in the right-hand side of Eq.~\eqref{eq:incompatibilityFi} represents the total damage within the region $\Omega$ while the second term represents the effect of the interface. Now following Gupta and Steigmann~(2012)~\cite{gupta2012plastic}, the interface damage density at $\kappa_R({\mathcal{B}})$ may be defined as
\begin{equation}
    \mathbf{\Sigma}^T\,\tilde{\mathbf{t}}_1\,=\llbracket{\mathbf{F}\,^i}\rrbracket\, (\tilde{\mathbf{t}}_1\,\times\,\tilde{\mathbf{{N}}}).
\end{equation}
The geometric interpretation of the jump condition along the interface is also available in Vignolo~\textit{et al.}~(2018, 2019)~\cite{vignolo2018junction,vignolo2019some}.

\section{A kinematic framework for laminated composites undergoing dissipative processes}\label{sec:Three-term decomposition in composites}

We now use the aforementioned framework to analyze the mechanics of fiber-reinforced laminated composites undergoing mechanical damage. Although the multiple natural configurations framework is suitable for this purpose, the framework was primarily developed for a single-phase material undergoing a wide class of dissipative processes. To capture the behavior of multi-phase materials such as laminated composites, this framework requires certain modifications. Recently, Wijaya~\textit{et al.}~(2025)~\cite{wijaya2025mixture} used this framework to model thermochemical curing in laminated composites. Apart from the basic framework, the proposed model is quite different from Wijaya~\textit{et al.}~(2025)~\cite{wijaya2025mixture} since it is developed particularly for the current purpose of characterization of mechanical damage. 

Let us consider a body $\mathcal{B}_0$, made up of fiber-reinforced laminated composites. Following Bedford and Stern~(1972)~\cite{bedford1972multi}, we consider that any material particle in the body $\mathcal{B}_0$ is a \emph{composite particle} $P$ which consists of two constituent phases: matrix $(m)$ and fiber $(f)$. As mentioned in~\S~\ref{sec:Kinematics}, a motion $\chi(\mathbf{X},t)$ maps the composite material particle from the undeformed configuration $\kappa_R(\mathcal{B}_0)$ of the body to its current configuration $\kappa_t(\mathcal{B}_0)$. Let us now apply an instantaneous elastic unloading (i.e., all the external stimuli are removed) of an infinitesimal neighborhood around the composite material particles~\cite{rajagopal1998mechanics}. The instantaneous elastic unloading, denoted by $\mathbf{F}^{e^{-1}}$ takes the composite particles from its current configuration to a locally stress-free natural configuration, $\kappa_i(\mathcal{B}_0)$. So far, the configurations and the tangent maps follow the traditional theory of multiple natural configurations with an important distinction that the material particle under consideration is a composite particle. 
\begin{figure}[h]
    \centering
    \includegraphics[width=0.8\linewidth]{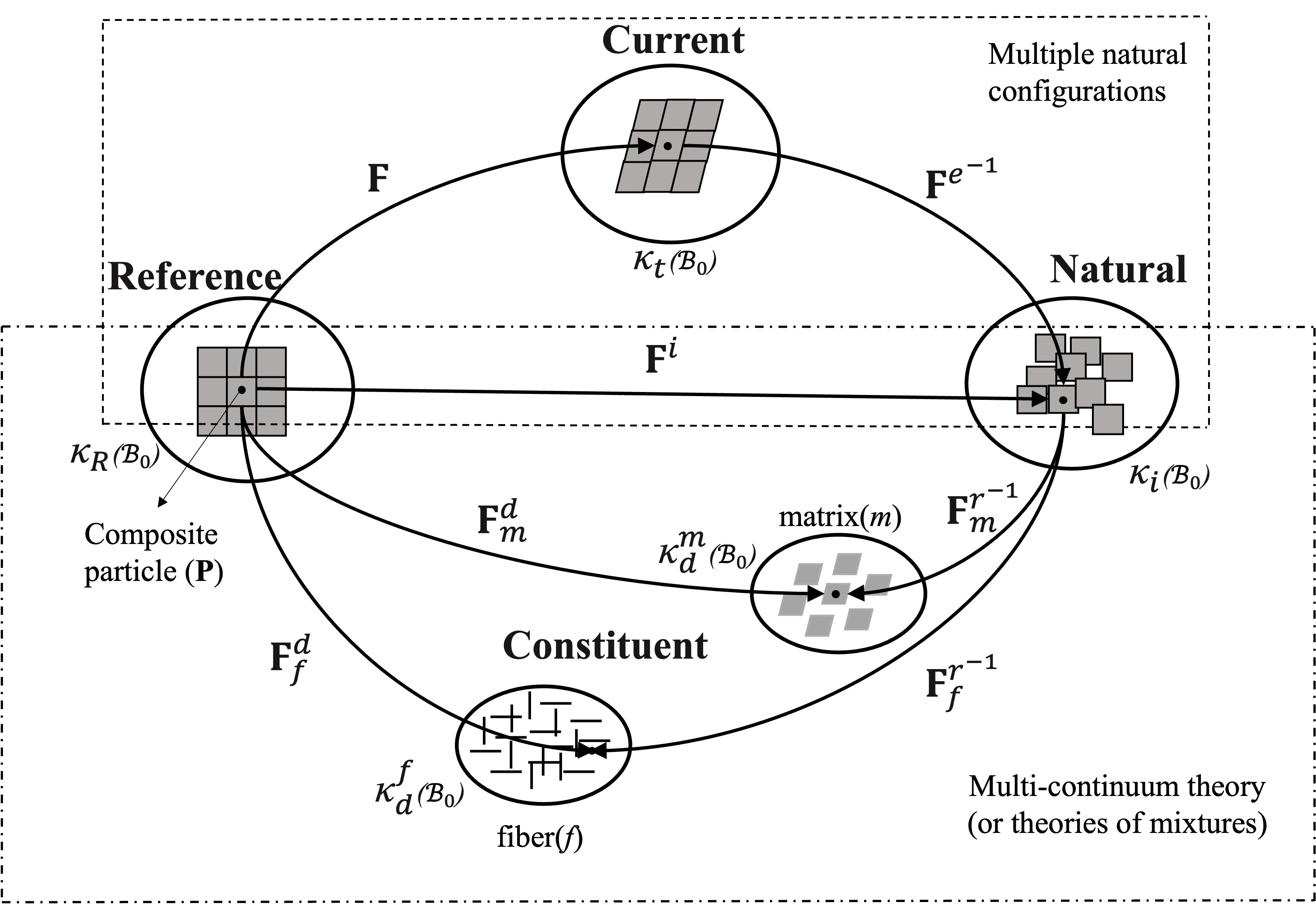}
    \caption{The configurations describing the deformation of a lamina of a fiber-reinforced composite material undergoing damage and the associated tangent maps.}
    \label{fig:three term}
\end{figure}

After removing all the external stimuli through the instantaneous elastic unloading, there still exist interaction forces between the constituents within a particular composite particle in the natural configuration $\kappa_i(\mathcal{B}_0)$. This interaction force also appears in the local balance law of the constituent. For example, for a particular constituent $\alpha$, the local form of the balance of momentum equation is written as~\cite{bedford1972multi}
\begin{equation}\label{eq:constituent_balance_law}
   {\text{div}}~\mathbf{t}_\alpha+\rho_\alpha(\mathbf{f}_\alpha-\dot{\mathbf{v}}_\alpha)+\rho_\alpha\,\mathbf{{r}}_\alpha=0.
\end{equation}
where $\rho$ is the partial density, $\mathbf{f}$ is the external force, $\mathbf{t}$ is the partial stress tensor and $\mathbf{r}_{\alpha}$ represents the interaction force on the constituent $\alpha$ exerted by the other constituent. This interaction force results from the consideration that the laminated composite is a superimposition of closely coupled interacting continua. Thus, a material particle in a laminated composite is co-occupied by its constituent continua.

For the purpose of characterizing different damage mechanisms in a laminated composite, we now employ the central idea of a multiple natural configurations framework to each composite material particle in the natural configuration of the body. For a composite particle, the interaction force is further removed in the same way as the instantaneous elastic unloading. The resulting constituents now constitute two different configurations, each consisting of a specific phase (i.e., either matrix or fiber) as shown in Fig.~\ref{fig:three term}. Let us denote these configurations by $\kappa^d_{\alpha}(\mathcal{B}_0)$, $\alpha$ can be either matrix ($m$) or fiber ($f$). It is important to note here that the configurations $\kappa^d_{\alpha}(\mathcal{B}_0)$ are not \emph{natural} in the sense of Rajagopal and Srinivasa~(1998)~\cite{rajagopal1998mechanics} where only external stimuli need to be removed. These configurations are an easy way to manifest the multi-continuum theory in terms of configurations and the related tangent maps. The tangent maps $\mathbf{F}^{r^{-1}}_{\alpha}$ maps an infinitesimal fiber belonging to a specific phase (matrix or fiber) of the composite particle in the tangent space of the natural configuration $\kappa_i(\mathcal{B}_0)$ to that of a different configuration. In this way, the process of removing the interaction forces for all the composite particles now results in two different configurations $\kappa^d_{\alpha}(\mathcal{B}_0)$ corresponding to the matrix and the fiber as shown in Fig.~\ref{fig:three term}. The material particles in $\kappa^d_{\alpha}(\mathcal{B}_0)$ are related to the corresponding composite particle in the undeformed configuration of the body through the tangent maps $\mathbf{F}\,^d_{\alpha}$. Therefore, for a particular composite particle, the tangent map $\mathbf{F}^i$ can be split in two different ways as
\begin{equation}\label{eq:constitutent_def_grad}
\mathbf{F}^i=\mathbf{F}^r_\alpha\,\mathbf{F}^d_\alpha \quad\text{where}\quad \alpha=f,~m.
\end{equation} 
This framework is consistent with the theory of interacting continua (mixture) and can also be found in constitutive modeling of viscoelastic materials, such as a generalized Maxwell solid~\cite{bonet2001large} and a Burgers fluid~\cite{malek2018derivation} etc. The two different decompositions of $\mathbf{F}^i$ stems from the idea that in a mixture, two different phases co-occupy the same material particle. The implication of this idea within the theory of multiple natural configurations has been expounded by M\'alek~\textit{et al.}~(2018)~\cite{malek2018derivation}.

Based on the kinematics discussed above, we now define the relevant Cauchy-Green tensors and the strain tensors corresponding to each tangent map. The right and the left Cauchy-Green tensors are defined as:
\begin{subequations}
\begin{equation}
\mathbf{C}^e=\mathbf{F}^{e^T}\,\mathbf{F}^e,\quad\mathbf{C}^r_\alpha=\mathbf{F}^{r^T}_\alpha\,\mathbf{F}^r_\alpha,\quad\mathbf{C}^d_\alpha=\mathbf{F}^{d^T}_\alpha\,\mathbf{F}^d_\alpha
\end{equation}
\text{and,}
\begin{equation}
\mathbf{B}^e=\mathbf{F}^{e}\,\mathbf{F}^{e^T},\quad \mathbf{B}^r_\alpha=\mathbf{F}^r_\alpha\,\mathbf{F}^{r^T}_\alpha,\quad\mathbf{B}^d_\alpha=\mathbf{F}^{d}_\alpha\,\mathbf{F}^{d^T}_\alpha.
\end{equation}
\end{subequations}
The Cauchy-Green tensors act as metrics in the corresponding configurations of the body. Now, the Green strain tensor and the Almansi-Hamel strain tensors corresponding to the associated tangent maps can be defined as
\begin{subequations}
\begin{equation}\label{eq:3_green_strain}
     \mathbf{E}^e=\dfrac{1}{2}\left(\mathbf{C}^e-\mathbf{I}\right),\quad\mathbf{E}^d_\alpha=\dfrac{1}{2}\left(\mathbf{C}^d_\alpha-\mathbf{I}\right),\quad\mathbf{E}^r_\alpha=\dfrac{1}{2}\left(\mathbf{C}^r_\alpha-\mathbf{I}\right),
\end{equation}
\text{and,}
\begin{equation}\label{eq:Almansi_green_strain}
     \mathbf{E}^{e^*}=\dfrac{1}{2}\left(\mathbf{I}-\mathbf{B}^{e^{-1}}\right),\quad\mathbf{E}^{r^*}_\alpha=\dfrac{1}{2}\left(\mathbf{I}-\mathbf{B}^{r^{-1}}_\alpha\right),\quad\mathbf{E}^{d^*}_\alpha=\dfrac{1}{2}\left(\mathbf{I}-\mathbf{B}^{d^{-1}}_\alpha\right).
\end{equation}
\end{subequations}
Our proposed kinematic framework for fiber-reinforced laminated composites is now ready for their applications in the characterization of damage.
 
\section{Characterization of damage}\label{Damage characterization}

\begin{figure}[h]
    \centering
    \includegraphics[width=0.8\linewidth]{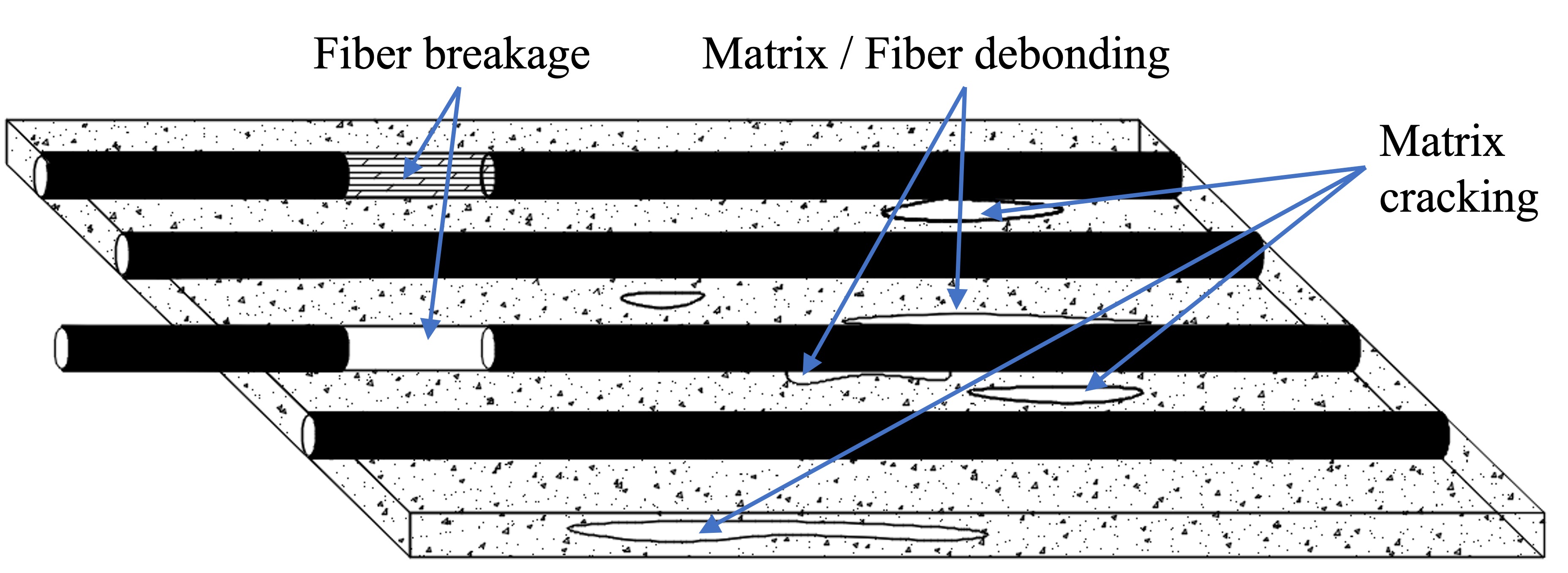}
    \caption{Different damage mechanisms in a lamina of a fiber-reinforced composite material. The damage mechanisms shown here are fiber breakage, matrix cracking, and interfacial debonding.}
    \label{fig:damage}
\end{figure}

The mechanical damage in composite materials is distinct from that of single-phase materials owing to the different damage mechanisms that these materials exhibit. The primary damage mechanisms considered here are fiber breakage, matrix cracking, interfacial debonding, and delamination as shown in Fig.~\ref{fig:damage}. This section deals with the characterization of these damage mechanisms. In general, there are two prevalent methods for (theoretical) characterization of material defects. The material defects can be quantified from a geometric perspective by analyzing the metric incompatibility of the associated configurations. This metric incompatibility can be expressed in terms of a non-vanishing torsion and/or curvature. Another popular approach, as discussed in \S~\ref{sec:Preliminaries}, is the physical arguments in terms of closure failure of a path circuit. These two methods often yield similar results for the quantification of material defects. In this section, we used the latter to quantify damage in laminated composites.

\subsection{Matrix cracking and fiber breakage}\label{sec:matrix}
Let us consider an initially closed circuit in the undeformed configuration $\kappa_R(\mathcal{B}_0)$ of the body made up of composite particles as shown in \S~\ref{sec:Continuous and discontinuous manifold}. In this configuration, the circuit is closed since we assumed the initial undeformed configuration to be damage-free. Hence, the path integral calculated in this flat Euclidean space vanishes. On the other hand, when the path integral along the circuit is calculated in the constituent configurations $\kappa^d_\alpha(\mathcal{B}_0)$, the path integral yields a non-zero value, implying that the circuit is no longer closed. The non-vanishing path integral can be attributed to the existence of corresponding damage contents. In this section, we are interested in finding the damage content specific to a particular phase, viz., the matrix and the fiber. For this purpose, two different approaches can be taken. The first approach involves computing the path integral along a circuit in the constituent configuration $\kappa^d_{\alpha}(\mathcal{B}_0)$, which was initially defined on the undeformed configuration of the body. The other approach would be to express the cumulative damage content in terms of the current configuration of the body. Needless to say, the former definition is based on a Lagrangian formulation, whereas the latter is its equivalent counterpart in an Eulerian framework. These approaches are similar to the one described in \S~\ref{sec:Continuous and discontinuous manifold} as well as the ones prescribed by Cermelli and Gurtin~(2001)~\cite{cermelli2001characterization} in the context of plasticity. For the Eulerian formulation, since the configuration $\kappa^d_\alpha(\mathcal{B}_0)$ is obtained by applying $\mathbf{F}^{r^{-1}}_\alpha$ to the material particles in the natural configuration, one may want to define the initial circuit in the natural configuration $\kappa_i(\mathcal{B}_0)$. However, as the natural configuration ${\kappa}_i(\mathcal{B}_0)$, is not Euclidean, one needs to define the initial circuit in the current configuration. Let us perform these exercises first for the matrix constituent configuration, i.e., $\mathbf{F}^i=\mathbf{F}^r_m\,\mathbf{F}^d_m$. 

\begin{figure}[h]
	\centering
	\begin{subfigure}{0.45\linewidth}
		\includegraphics[width=0.9\linewidth]{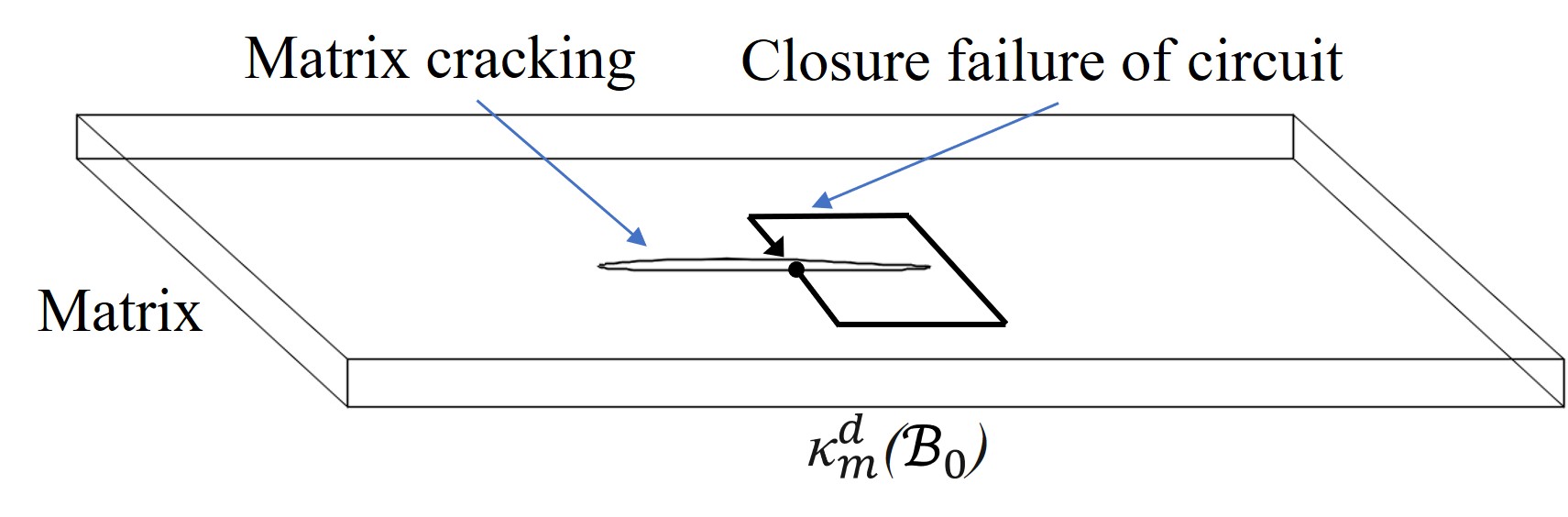}
		\caption{}
		\label{fig:matrix_crack}
	\end{subfigure}
	\begin{subfigure}{0.45\linewidth}
		\includegraphics[width=0.9\linewidth]{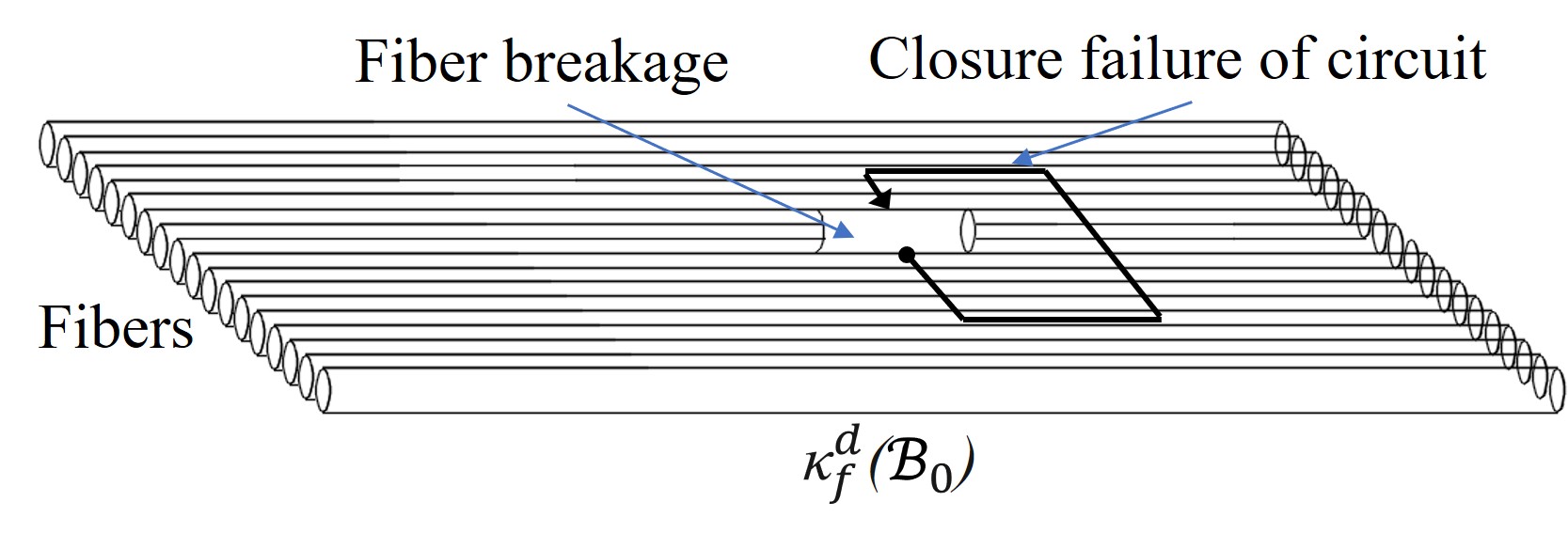}
		\caption{}
		\label{fig:fiber_breakage}
	\end{subfigure}
    \caption{Schematic diagram of material defects and respective circuit: (a) Matrix cracking and the circuit in $\kappa^d_m(\mathcal{B}_0)$ and (b) Fiber breakage and the circuit in $\kappa^d_f(\mathcal{B}_0)$.}
	\label{fig:matrix_crack_fiber_breakage}
    \end{figure}

For matrix cracking, let us take an initially closed circuit in the reference configuration $\kappa_R(\mathcal{B}_0)$. The circuit in the matrix constituent configuration after deformation is shown in Fig.~\ref{fig:matrix_crack}. Since an infinitesimal fiber $d\mathbf{X}$ is related to the matrix constituent configuration $\kappa^d_m(\mathcal{B}_0)$ via the tangent map $\mathbf{F}^d_m$, a path integral along a curve $\partial\Omega$ is evaluated as 
\begin{equation}\label{eq:burgers_vector_reference}
    \mathbf{b}^R_{m}=\int_{\partial\Omega} \mathbf{F}^{d}_m\,d\mathbf{X}=\int_\Omega(\text{Curl}\,\mathbf{F}^d_m)^T\,\mathbf{N}\,d{A}.
\end{equation}
The second of Eq.~\eqref{eq:burgers_vector_reference} is obtained by using Stokes' theorem. $\mathbf{b}^R_m$ provides a measure of incompatibility of the configuration $\kappa^d_m(\mathcal{B}_0)$ measured from the undeformed configuration of the body. Physically, this incompatibility represents the accumulated matrix cracks in the surface enclosed by $\partial\Omega$ following the definition of Kachanov~(1980)~\cite{kachanov1980continuum}. The measure of incompatibility is further transformed by pushing forward the vector area from the reference configuration $\kappa(\mathcal{B}_0)$  of the body into its matrix constituent configuration. The resulting damage content is written as 
\begin{equation}\label{eq:burgers_vector_matrix_configuration_Pushing}
\mathbf{b}^R_{m}=\int_\Omega \dfrac{1}{J^d_m}(\text{Curl}\,\mathbf{F}^d_m)^T\,\mathbf{F}^{d^T}_m\,\mathbf{n}_m\,d{a_m}.
\end{equation}
Here, $\mathbf{n}_m\,da_m$ is the infinitesimal vector area in the matrix constituent configuration $\kappa^d_m(\mathcal{B}_0)$ of the body and the Jacobian of the map is given as $J^d_m = \text{det}(\mathbf{F}^d_m)$. The damage content can also be written from the current configuration of the body. For this purpose, we first consider the initially closed curve in the current configuration. Now following the same exercise, the damage content can be written as
\begin{equation}\label{eq:burgus_vector_matrix_current}
     \mathbf{b}^t_{m}=\int_{\partial\omega} (\mathbf{F}^{e^{-1}}\, \mathbf{F}^{r^{-1}}_m)\,d\mathbf{x}=\int_\omega(\text{curl}\,(\, \mathbf{F}^{e^{-1}}\, \mathbf{F}^{r^{-1}}_m))^T\,{\mathbf{n}}\,da.
\end{equation}
Here, $\partial\omega$ is a closed path in the current configuration of the body. Now transforming $\mathbf{b}^t_m$ into the matrix configuration $\kappa^d_m(\mathcal{B}_0)$ by using a pull-back operation, $\mathbf{b}^t_m$ can be written as 
\begin{equation}\label{eq:burgus_vector_matrix_current_pullback}
    \mathbf{b}^t_{m}= \int_\omega\, J^e\,J^r_m(\text{curl}\,(\mathbf{F}^{e^{-1}}\,\mathbf{F}^{r^{-1}}_m))^T\,\mathbf{F}^{r^{-T}}_m\,\mathbf{F}^{e^{-T}}\,{\mathbf{n}_m}\,da_m.
\end{equation}
It is worth noting that the Eqs.~\eqref{eq:burgers_vector_matrix_configuration_Pushing} and \eqref{eq:burgus_vector_matrix_current_pullback} are both measures of the same incompatibility, one written from the reference configuration and the other from the current configuration. In the spirit of crack density tensor, one can now define a second-order tensor $\mathbf{G}_m$ such that $\mathbf{G}^T_m\,\mathbf{n}_m~da_m$ represents the cumulative matrix (micro) crack inside a region enclosed by the circuit in the matrix configuration, $\kappa^d_m(\mathcal{B}_0)$. Therefore, $\mathbf{G}_m$ can be viewed as a density of continuously distributed matrix cracks in that configuration. The matrix crack density can be written as
\begin{equation}\label{eq:matrix_crack_density}
    \mathbf{G}_m=\dfrac{1}{J^d_m}\,\mathbf{F}^{d}_m\,(\text{Curl}\,\mathbf{F}^d_m)=J^e\,J^r_m\,\mathbf{F}^{e^{-1}}\,\mathbf{F}^{r^{-1}}_m\,\text{curl}\,(\mathbf{F}^{e^{-1}}\,\mathbf{F}^{r^{-1}}_m).
\end{equation}

Similarly, for the fiber breakage {shown in Fig.~\ref{fig:fiber_breakage}}, the damage content and its density can be obtained following the same procedure for the configuration $\kappa^d_f(\mathcal{B}_0)$. In this case, the measure of incompatibility can be written as  
\begin{equation}\label{eq:burgers_vector_reference_fiber}
    \mathbf{b}_{f}=\int_{\partial\Omega} \mathbf{F}^{d}_f\,d\mathbf{X}=\int_\Omega(\text{Curl}\,\mathbf{F}^d_f)^T\,\mathbf{F}^{d^T}_f\,\mathbf{n}_f\,d{a}_f=\int_{\omega}J^e\,J^r_f(\text{curl}\,(\mathbf{F}^{e^{-1}}\,\mathbf{F}^{r^{-1}}_f))^T\,\mathbf{F}^{r^{-T}}_f\,\mathbf{F}^{e^{-T}}\,{\mathbf{n}_f}\,da_f.
\end{equation}
The corresponding density of fiber breakage assuming a continuous distribution of broken fibers in the configuration $\kappa^d_f(\mathcal{B}_0)$ can be written as
\begin{equation}\label{eq:burgus_vector_fiber_current}
     \mathbf{G}_f=\dfrac{1}{J^d_f}\,\mathbf{F}^{d}_f\,(\text{Curl}\,\mathbf{F}^d_f)=J^e\,J^r_f\,\mathbf{F}^{e^{-1}}\,\mathbf{F}^{r^{-1}}_f\,\text{curl}\,(\mathbf{F}^{e^{-1}}\,\mathbf{F}^{r^{-1}}_f).
\end{equation}

\subsection{Debonding and interfacial slip}\label{sec:debonding}

In our study of mechanical damage in laminated composites, so far we have dealt with only one particular phase of a lamina, i.e., either a matrix or a fiber. We now characterize the damage mechanism, namely interfacial slip and debonding, that involves both these phases within a particular lamina. For this purpose, we study the interfacial slip between the two constituents in terms of their relative distortion tensors, akin to the work of Cermelli and Gurtin~(1994)~\cite{cermelli1994kinematics} for a multiphase material undergoing phase transition. Before going into the damage characterization for a composite lamina, let us first briefly revisit this framework. Let us consider a multiphase material with phases $\mu$ and $\nu$ undergoing a motion. Let {$\mathbf{X}_{\beta},\:\beta=\mu,\nu$} denote the position vectors of a material particle corresponding to the phases at the interface $S_{\beta}$ between them. The individual motions of these phases are given as 
\begin{equation}\label{eq:motion_Gurtin}
    \mathbf{x}_\beta=\mathcal{X}_{\beta}(\mathbf{X}_\beta(t),t),\quad \beta=\mu, \nu.
\end{equation}
If the two phases $\mu$ and $\nu$ are perfectly bonded to each other, then their motions must lead to the same material particle in the current configuration of the body and thus, $\mathcal{X}_{\mu}(\mathbf{X}_\mu,t)=\mathcal{X}_{\nu}(\mathbf{X}_\nu,t)$. Now, a convected material derivative (Lie derivative) of this condition leads to
\begin{equation}\label{eq:motion_slip_1}
    (\dot{\mathcal{X}})_\mu+\mathbf{F}_\mu \dot{\mathbf{X}}_\mu=(\dot{\mathcal{X}})_\nu+\mathbf{F}_\nu \dot{\mathbf{X}}_\nu.
\end{equation}
Eq.~\eqref{eq:motion_slip_1} implies that the relative velocity between the two phases is zero. Now, if the phases are not perfectly bonded, the interfacial slip between the two phases can be measured in terms of the relative velocity of the individual motions, given by
\begin{equation}\label{eq:gamma_def}
    \boldsymbol{\gamma}=(\dot{\mathcal{X}})_\mu-(\dot{\mathcal{X}})_\nu+\mathbf{F}_\mu \dot{\mathbf{X}}_\mu-\mathbf{F}_\nu \dot{\mathbf{X}}_\nu.
\end{equation}
Cermelli and Gurtin~(1994)~\cite{cermelli1994kinematics} defined the interfacial slip rate as the difference between the tangential components of the relative velocity along the interface in the current configuration of the body. Thus, the interfacial slip rate can be written as 
\begin{equation}\label{eq:gamma_tangent}
    \boldsymbol{\lambda}_t=\mathbf{F}_\mu\,(\dot{\mathbf{X}}_\mu)_{\parallel}-\mathbf{F}_\nu\,(\dot{\mathbf{X}}_\nu)_{\parallel}\defeq\llbracket{\mathbf{F}{(\dot{\mathbf{X}})}_{\parallel}}\rrbracket.
\end{equation}
{This idea is particularly useful for the characterization of interfacial debonding and slip in laminated composites. Due to the differential deformation in the matrix and the fiber under the application of load, the developed shear stress along their interface exceeds the interfacial shear strength, causing debonding between the two phases~\cite{hsueh1990interfacial,talreja2012damage}. It is evident that the shear stress developed at the interface is a thermodynamic conjugate to the relative tangential velocity along the interface between the matrix and the fiber. Thus, from a kinematic point of view, this relative tangential velocity characterizes debonding and interfacial slip.} However, we notice that a relation between the velocities corresponding to the two phases in Eq.~\eqref{eq:motion_slip_1} cannot be directly written for our framework due to the lack of a global compatibility of the configurations $\kappa^d_\alpha(\mathcal{B}_0)$. Therefore, the definition~\eqref{eq:gamma_def} will not be useful in our case. We shall use a variant of Eq.~\eqref{eq:gamma_tangent} as a measure of interfacial slip and debonding for laminated composites.

We now employ this idea to characterize the damage content resulting from debonding and interfacial slip in a laminated composite material. In our framework, the phases $\mu$ and $\nu$ can be identified with the two constituents of a laminated composite, i.e, the matrix $(m)$ and the fiber $(f)$. A significant difference between the framework of Cermelli and Gurtin~(1994)~\cite{cermelli1994kinematics} and our kinematics is that we consider a motion of a \emph{composite} particle consisting of the matrix and the fiber instead of their individual motions separately. Therefore, the relative motion between the matrix and the fiber can only be characterized in terms of the tangent map between the reference $\kappa_R(\mathcal{B}_0)$ and the constituent configurations $\kappa^d_\alpha(\mathcal{B}_0)$. For a general inhomogeneous deformation, the tangent map $\mathbf{F}^d_{\alpha}$ is not integrable and thus, it cannot be written as a gradient of a motion as in Eq.~\eqref{eq:motion_Gurtin}. As mentioned by Rajagopal and Srinivasa~(1998)~\cite{rajagopal1998mechanics}, this quandary can be resolved by either considering a fictitious local homogeneous motion such that the gradient of this motion is equivalent to the tangent maps $\mathbf{F}^d_\alpha$ or by using sophisticated tools from differential geometry considering $\kappa^d_{\alpha}(\mathcal{B}_0)$ as non-Euclidean spaces. To keep our analysis tractable, we consider the former approach in this paper. 

Let us consider a fictitious, locally homogeneous motion $\boldsymbol{\zeta}_{\alpha}$ from the reference configuration of the body to the constituent configurations $\kappa^d_{\alpha}(\mathcal{B}_0)$ such that the gradient of $\boldsymbol{\zeta}_{\alpha}$ produces a second-order tensor equivalent to the tangent maps $\mathbf{F}^d_{\alpha}$ corresponding to the matrix and the fiber. Since the matrix and the fiber co-occupy the same composite particle in the reference configuration of the body with position vector $\mathbf{X}(t)$, their individual local motions can be written as
\begin{equation}\label{eq:individual_motion_mf}
    \mathbf{x}^d_{\alpha}=\boldsymbol{\zeta}_{\alpha}(\mathbf{X}(t),t)\quad \alpha=m,f.
\end{equation}
The local spatial velocities of the constituents can be written as a time derivative of the associated motions, viz.,
\begin{equation}\label{eq:individual_vel}
\mathbf{v}^d_{\alpha}=\dfrac{\partial\boldsymbol{\zeta}_{\alpha}}{\partial t},\quad \alpha=m,f.
\end{equation}
Following Cermelli and Gurtin~(1994)~\cite{cermelli1994kinematics}, the interfacial slip can now be defined as a difference between the tangential components of these velocities akin to Eq.~\eqref{eq:gamma_tangent}. This definition, however, cannot be directly used since the material particles corresponding to a specific phase such as a matrix or a fiber, that belong to a composite particle $\mathbf{X}$ in $\kappa_R(\mathcal{B}_0)$, is mapped to two different configurations $\kappa^d_m(\mathcal{B}_0)$ and $\kappa^d_f(\mathcal{B}_0)$ through the tangent maps $\mathbf{F}^d_m$ and $\mathbf{F}^d_f$, respectively. Therefore, the local velocities are measured in two different constituent configurations, and hence, they are vectors in the tangent space of $\kappa^d_{\alpha}(\mathcal{B}_0)$. To overcome this issue, we first pull back these velocities into the reference configuration $\kappa_R(\mathcal{B}_0)$ through~\cite{marsden1994mathematical}
\begin{equation}\label{eq:vel_pulled_back}
\mathbf{V}_{\alpha}=\mathbf{F}^{d^{-1}}_{\alpha}\,\mathbf{v}_{\alpha}.
\end{equation}
The relative velocity can now be defined as a difference between the spatial velocities, $\mathbf{v}_{\alpha}$ pulled back into the reference configuration of the body. The definition of the relative velocity reads 
\begin{equation}\label{eq:relative_vel_reference}
    \mathbf{V}_{\text{rel}}={\mathbf{V}}_m-{\mathbf{V}}_f=\mathbf{F}^{d^{-1}}_m\mathbf{v}^d_m\,-\mathbf{F}^{d^{-1}}_f\mathbf{v}^d_f.
\end{equation}
In a multi-phase flow problem, considered by Cermelli and Gurtin~(1994)~\cite{cermelli1994dynamics}, the interfacial slip rate was calculated in the current configuration of the body since the material particles from different configurations (and hence, different phases) are mapped into a single particle in the current configuration of the body. Our kinematic framework is opposite to this problem, as material particles belonging to a single composite particle in the reference configuration of the body, are mapped into two different constituent configurations. Therefore, we ought to consider a Lagrangian framework for our problem. Let us consider that an interface $S$ exists between the matrix and the fiber within a particular composite particle in the reference configuration of the body. Let $\mathbf{N}$ denote the normal to the interface $S$ in $\kappa_R(\mathcal{B}_0)$. The interfacial slip rate can now be defined in the reference configuration of the body as
\begin{equation}\label{eq:interfacial_slip_rate}
    \boldsymbol{\Lambda}=\overline{\mathbf{V}}_m-\overline{\mathbf{V}}_f\quad\text{where}\quad \overline{\mathbf{V}}_{\alpha}=\mathbf{V}_{\alpha}-(\mathbf{V}_{\alpha}\cdot\mathbf{N})\,\mathbf{N}.
\end{equation}
Since $\overline{\mathbf{V}}_\alpha$ are the tangential components of $\mathbf{V}_\alpha$, the definition of $\boldsymbol{\Lambda}$ is equivalent to Eq.~\eqref{eq:gamma_tangent}. {We assume that the contact between the two phases remain intact along the matrix-fiber interface during the process of interfacial slip.} Hence, the normal components of the pulled-back spatial velocities must be the same, and they must satisfy the condition\footnote{{Here we assume that the interfacial slip is primarily governed by the developed shear stress at the matrix-fiber interface and thus, the contact between the two phases remain intact~\cite{hsueh1990interfacial,talreja1991continuum}. Although a matrix-fiber separation along the normal to the interface is possible especially for matrix crack-induced debonding and interfacial sliding, such phenomena can be taken into account through the measure of incompatibility in Eq.~\eqref{eq:matrix_crack_density} at a point near the interface whose (infinitesimal) neighborhood lies completely in the matrix phase. Note that the coupling between the different damage mechanisms enters into our framework through the constitutive modeling as shown in \S~\ref{sec:constitutive}.}}
\begin{equation}\label{eq:vel_normal}
    (\mathbf{V}_{m}\cdot\mathbf{N})\,\mathbf{N}=(\mathbf{V}_{f}\cdot\mathbf{N})\,\mathbf{N}.
\end{equation}
In view of Eq.~\eqref{eq:vel_normal}, one can notice that the interfacial slip rate is numerically the same as the relative velocity, $\mathbf{V}_\text{rel}$. The interfacial slip rate can now be pushed forward into the matrix configuration as
\begin{equation}\label{eq:interfacial_slip_matrix}
    \boldsymbol{\lambda}_m=\mathbf{F}^d_m\,\mathbf{V}_{rel}=\mathbf{v}^d_m\,-\mathbf{F}^d_m\mathbf{F}^{d^{-1}}_f\mathbf{v}^d_f.
\end{equation}
Let us now introduce a relative distortion tensor $\mathbf{M}$ in terms of the tangent maps $\mathbf{F}^d_{\alpha}$ as
\begin{equation}\label{eq:relative_distorsion_tensor_Interface}
    {\mathbf{M}}\defeq{{\mathbf{F}}^{d}_m}\,{\mathbf{F}}{^{d^{-1}}_f}. 
\end{equation} 
Using this relative distortion tensor, Eq.~\eqref{eq:interfacial_slip_matrix} can be written as
\begin{equation}\label{eq:relative_vel_matrix}
    \boldsymbol{\lambda}_{m}=\mathbf{v}^d_m\,-\mathbf{M}\,\mathbf{v}^d_f.
\end{equation}
The relative distortion tensor $\mathbf{M}:T\boldsymbol{\zeta}_f(\mathcal{B}_0)\rightarrow T\boldsymbol{\zeta}_m(\mathcal{B}_0)$ is a two-point tensor that takes any vector from the tangent space of the configuration $\kappa^d_f(\mathcal{B}_0)$ and places it into the tangent space of $\kappa^d_m(\mathcal{B}_0)$. Therefore, the interfacial slip rate in Eq.~\eqref{eq:relative_vel_matrix} represents the difference between the spatial velocities corresponding to the motions $\boldsymbol{\zeta}_{\alpha}$, now both written in the same configuration, $\kappa^d_m(\mathcal{B}_0)$. 

Needless to say that for a perfectly bonded lamina, the interfacial slip must be zero, i.e.,
\begin{equation}\label{eq:perfect_bond_vel}
    \boldsymbol{\lambda}_m=\mathbf{0}\implies \mathbf{v}^d_m\,=\mathbf{M}\,\mathbf{v}^d_f.
\end{equation}
This condition is rather counterintuitive as the no-slip condition not only depends on the velocities but also on the relative distortion tensor. Now let us consider a special subcase of Eq.~\eqref{eq:perfect_bond_vel} in which the matrix and the fiber corresponding to a particular composite particle occupy the same position $\mathbf{x}^d$ in the constituent configuration, i.e., $\boldsymbol{\zeta}_m(\mathbf{X},t)=\boldsymbol{\zeta}_f(\mathbf{X},t)$. Note that this condition does not ensure that the two configurations $\kappa^d_m(\mathcal{B}_0)$ and $\kappa^d_f(\mathcal{B}_0)$ coincide as $\boldsymbol{\zeta}_{\alpha}(\mathbf{X},t)$ is a \emph{local} fictitious motion corresponding to  $\mathbf{F}^d_{\alpha}$. Under this condition, it can be easily shown that the relative distortion tensor turns out to be a second-order identity tensor along with the condition that $\boldsymbol{\dot{\zeta}}_m=\boldsymbol{\dot{\zeta}}_f$. Now substituting these two conditions together in Eq.~\eqref{eq:relative_vel_matrix}, one can show that $\boldsymbol{\lambda}_m=\mathbf{0}$. Finally, the interfacial slip rate in the fiber configuration $\kappa^d_f(\mathcal{B}_0)$ can be written as
\begin{equation}
     \boldsymbol{\lambda}_{f}=\mathbf{M}^{-1}\mathbf{v}^d_m\,-\mathbf{v}^d_f.
\end{equation}
In view of the physical meaning of the relative distortion tensor $\mathbf{M}$, the interfacial slip rate $\boldsymbol{\lambda}^d_f$ is simply $\boldsymbol{\lambda}^d_m$ pulled back into the configuration $\kappa^d_f(\mathcal{B}_0)$ via
\begin{equation}
    \boldsymbol{\lambda}_f=\mathbf{M}^{-1}\,\boldsymbol{\lambda}_m.
\end{equation}

\subsection{Delamination}\label{sec:delamination}

For the characterization of damage studied so far, we have considered a single lamina made up of fibers and matrix. In this section, we extend this study further to characterize the damage content when more than one lamina is present. For the ease of the current study, we consider a laminate comprising of only two lamin\ae. In the undeformed configuration of the body $\kappa_R(\mathcal{B}_0)$, these lamin$\ae$ are perfectly bonded to each other, implying that there is no jump in the displacement field across the interface between them. When undergoing some dissipative process, these two lamin$\ae$ may separate at certain parts of the interface in the current configuration $\kappa_t(\mathcal{B}_0)$. Our objective is to characterize the content of delamination between the two lamin\ae~by studying the jump in the displacement field across the interface between them. The development in this section closely follows the formulation of Gupta and Steigmann~(2012)~\cite{gupta2012plastic}, although their work was carried out in the context of plasticity. It is worth noting that the interface {between two phases} played a significant role in our previous characterization of debonding and interfacial slip as well. {However, an interface between two lamin\ae,~each subjected to intra-laminar damage, needs to be considered for the characterization of delamination. The information regarding different intra-laminar damage corresponding to matrix cracks, fiber breakage and debonding in each lamina $\delta$ are encoded in the respective tensor fields $\mathbf{F}^i_{\delta}$ since $\mathbf{F}^i_{\delta}=\mathbf{F}^r_{\alpha_{\delta}}\,\mathbf{F}^d_{\alpha_{\delta}}, \alpha=m,f$. Hence, for delamination, it is sufficient to analyze the jump in the displacement field across the interface between two lamin\ae~in the natural configuration of the body $\kappa_i(\mathcal{B}_0)$, obtained through the decomposition $\mathbf{F}=\mathbf{F}^e\,\mathbf{F}^i$ in accordance with the theory of multiple natural configurations.}
\begin{figure}[h]
    \centering
    \includegraphics[width=0.8\linewidth]{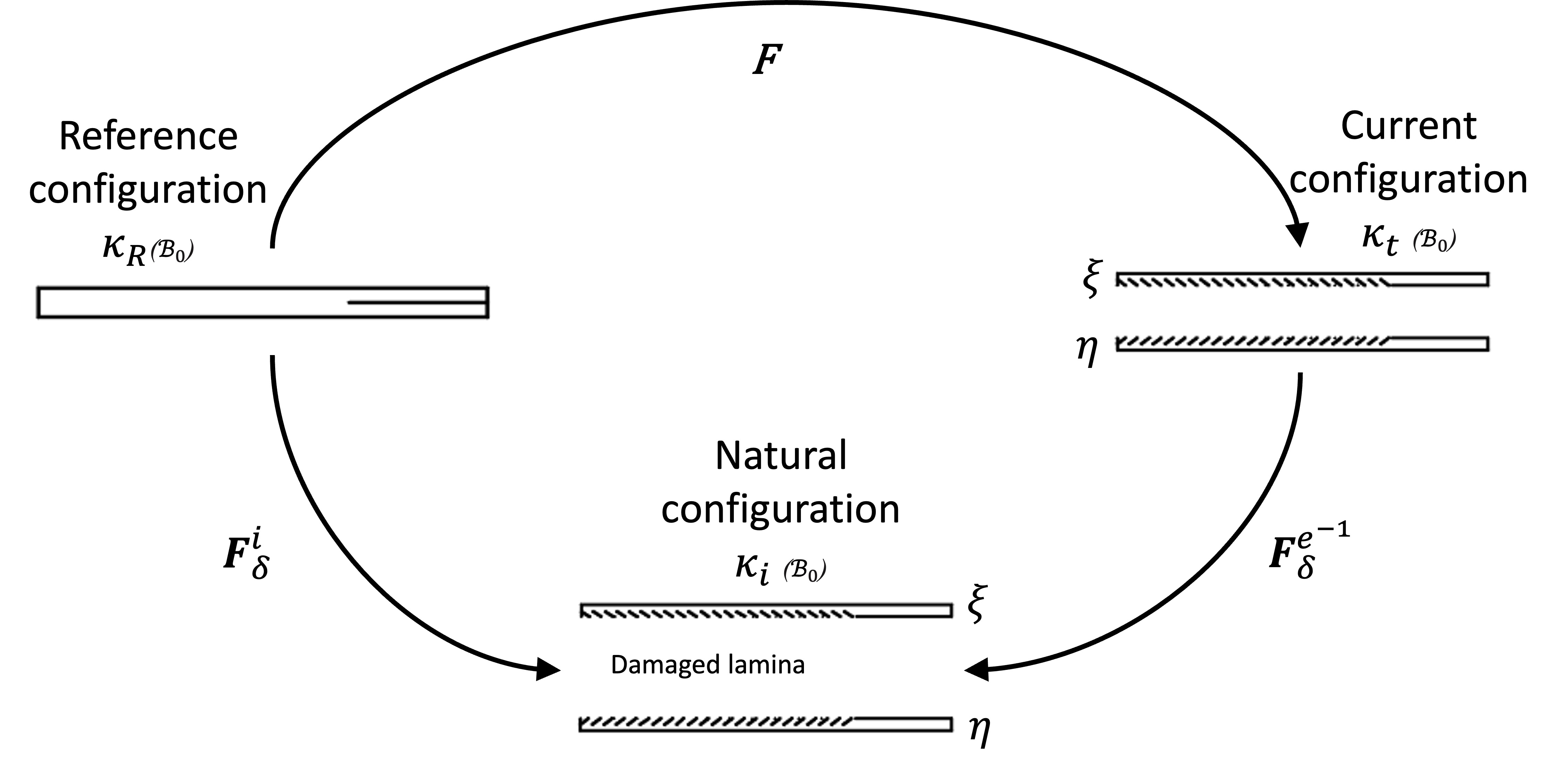}
    \caption{Configurations associated with delamination and the respective tangent maps.}
    \label{fig:delamination}
\end{figure}

Now let us consider a body $\mathcal{B}_0$ made up of two lamin$\ae$, denoted by $\xi$ and $\eta$ as shown in Fig.~\ref{fig:delamination}. The body follows the same kinematics as described in \S~\ref{sec:Kinematics}. Although the tangent maps $\mathbf{F}^e$ and $\mathbf{F}^i$ map the tangent vectors at a single material particle, here we have used the notation $\mathbf{F}^e_{\delta}$ and $\mathbf{F}^i_{\delta},\,\text{where}\, \delta=\xi\, \text{or}\, \eta$, to account for the tangent maps at a material particle corresponding to the lamin$\ae$ $\xi$ and $\eta$, respectively. Let $S_R$ be the interface at the reference configuration between the two lamin$\ae$ and $\Gamma$ denote a curve at the intersection of the bulk parts of the lamin$\ae$, $\Omega_{\delta}$ such that $\Gamma=S_R\,\cap\,\Omega_{\delta}$. To measure the damage content, let us follow the same exercise as in \S~\ref{sec:Continuous and discontinuous manifold} in the natural configuration of the body. The total damage content in the perfectly bonded lamina, i.e., when there is no jump in the displacement field across the interface, can be obtained as 
\begin{equation}\label{eq:bulk_burgers}
     \mathbf{b}^R=\int_{\partial\Omega}\mathbf{F}^id\mathbf{X}.
\end{equation}
In the presence of delamination, it is reasonable to assume that $\mathbf{F}^i$ is continuous on the bulk parts of both the lamin$\ae$, $\Omega_{\delta}$. However, a jump in $\mathbf{F}^i$ can be observed across the interface. Under this consideration, Eq.~\eqref{eq:bulk_burgers} can be split into the bulk and the interface part as
\begin{equation}\label{eq:bulk_inteface_del}
    \int_{{{\partial\Omega}_{\xi}}\,\cup\,{{\partial\Omega}_{\eta}}}\mathbf{F}^i\,d\mathbf{X}= \int_{{{\Omega}_{\xi}}\,\cup\,{{\Omega}_{\eta}}}(\text{Curl}\,\mathbf{F}^i)^T\,\mathbf{N}\,dA= \int_{\partial\Omega}\mathbf{F}^id\mathbf{X}+\int_\Gamma\llbracket\mathbf{F}^i\rrbracket d\mathbf{X}.
\end{equation}
To understand the physical meaning of Eq.~\eqref{eq:bulk_inteface_del}, this equation is rewritten as
\begin{equation}\label{eq:Del_Burgers}
  \int_{\partial\Omega}\mathbf{F}^id\mathbf{X}=\int_{\Omega}(\text{Curl}\,\mathbf{F}^i)^T\,\mathbf{N}\,dA-\int_\Gamma\llbracket\mathbf{F}^i\rrbracket d\mathbf{X}.
\end{equation}
The first of the right-hand side of Eq.~\eqref{eq:Del_Burgers} represents the total damage content in the bulk part of the lamin$\ae$ $\xi$ and $\eta$, whereas the second part accounts for the damage content associated with their interface. The latter is of particular interest as it represents the delamination between the two lamin$\ae$. The rest of the section is devoted to further understanding of this part.

Now let us define the a set of orthogonal bases $\tilde{\mathbf{t}}_1$ and $\tilde{\mathbf{t}}_2$ in the tangent space of the interface $S_R$ in $\kappa_R(\mathcal{B}_0)$. Let $\tilde{\mathbf{N}}$ denote the normal to the interface $S_R$ such that $\tilde{\mathbf{N}}=\tilde{\mathbf{t}}_1\times \tilde{\mathbf{t}}_2$ as shown in Fig.~\ref{fig:Interface}. The density of damage content due to delamination $\boldsymbol{\Sigma}$, corresponding to the interface in the reference configuration, is defined as
\begin{equation}\label{eq:tangential_mismatch}
    \mathbf{\Sigma}^T\tilde{\mathbf{t}}_1=\llbracket{\mathbf{F}\,^i}\rrbracket\,(\tilde{\mathbf{t}}_1\times\tilde{\mathbf{N}}).
\end{equation}
Now substituting Eq.~\eqref{eq:tangential_mismatch} into Eq.~\eqref{eq:Del_Burgers} (for detailed calculation, see~\ref{sec:appendix_del_total}), we get 
\begin{equation}\label{eq:burgers_interface_reference}
     \mathbf{b}^R= \underbrace{\int_{\Omega}(\text{Curl}\,\mathbf{F}^i)^T\,\mathbf{N}\,dA}_{\text{Bulk}}+\underbrace{\int_\Gamma  \mathbf{\Sigma}^T\tilde{\mathbf{t}}_1d{L}}_{\text{Interface}}.
\end{equation}
$\mathbf{b}^R$ in Eq.~\eqref{eq:burgers_interface_reference} is the total measure of incompatibility from the undeformed configuration of the body. Physically, this incompatibility represents the accumulation of the damage accounting for both the damage in the area bounded by $\partial\Omega$ and away from the interface (first term in right-hand side of Eq.~\eqref{eq:burgers_interface_reference}) and, damage at the interface, denoted by $\Gamma$. In a similar manner, the total damage content can also be expressed with respect to the current configuration as 
\begin{equation}\label{eq:burgers_interface_current} \mathbf{b}^t=\int_{\partial\omega}\mathbf{F}^{e^{-1}}d\mathbf{x}=\underbrace{\int_{\partial{\omega}}(\text{curl}\,\mathbf{F}^{e^{-1}})^T\,\mathbf{n}\,da}_{\text{Bulk}}+\underbrace{\int_\delta\boldsymbol{\sigma}^T\hat{\mathbf{t}}_1\,d{l}}_{\text{Interface}}
\end{equation} 
where $\hat{\mathbf{t}}_1$ is the base vector of the tangent space and $\boldsymbol{\sigma}$ is the damage density due to delamination at the interface in the current configuration of the body. Let us define an identity operator $\mathbf{1}\defeq\mathbf{N}\otimes\mathbf{N}+\tilde{\mathbf{t}}_1\otimes\tilde{\mathbf{t}}_1+\tilde{\mathbf{t}}_2\otimes\tilde{\mathbf{t}}_2$. Now using Eq.~\eqref{eq:tangential_mismatch} and the identity operator $\mathbf{1}$ on $\llbracket\mathbf{F}^i\rrbracket$ (cf.~\ref{sec:appendix_del_total}), we obtain
\begin{equation}\label{eq:nortmal_tangent_part}
   \mathbf{F}^i_\xi\,-\mathbf{F}^i_\eta= \mathbf{a}\otimes\tilde{\mathbf{N}}-\mathbf{\Sigma}^T\boldsymbol{\vartheta}\defeq \hat{\boldsymbol{\Sigma}}
\end{equation}
where {$\mathbf{a}$} is an arbitrary vector and $\boldsymbol{\vartheta}=\tilde{\mathbf{t}}_1\otimes\tilde{\mathbf{t}}_2-\tilde{\mathbf{t}}_2\otimes\tilde{\mathbf{t}}_1$. The first of the right-hand side of Eq.~\eqref{eq:nortmal_tangent_part} accounts for the normal part of $\llbracket\mathbf{F}^i\rrbracket$, whereas the last term accounts for its tangential part. {Note that the relative separation between the two lamin\ae~in the deformed configuration of the body are often used as a suitable kinematic description for delamination. For example, Zou~\textit{et al.}~(2001)~\cite{zou2001mode} considered the relative displacements between the corresponding points to compute the energy release rates for the different modes of delamination. In our framework, the position vectors of two points located at the lamin\ae~$\xi$ and $\eta$, that previously occupied the same position vector in the undeformed configuration of the body, is completely specified by the respective tangent maps $\mathbf{F}^i_{\delta}$ in the natural configuration $\kappa_i(\mathcal{B}_0)$. Hence, $\hat{\mathbf{\Sigma}}$ measures the relative separation of the two lamin$\ae$ and represents the damage density due to delamination.}

Now, to project $\llbracket\mathbf{F}^i\rrbracket$ on the two-dimensional interface, let us first define the projection tensor $\tilde{\mathbf{1}}$ such that $\tilde{\mathbf{1}}\defeq\tilde{\mathbf{t}}_1\otimes\tilde{\mathbf{t}}_1+\tilde{\mathbf{t}}_2\otimes\tilde{\mathbf{t}}_2$. Using the projection tensor on $\llbracket\mathbf{F}^i\rrbracket$, we arrive at
\begin{equation}\label{eq:jump_interface_projection} 
-\mathbf{\Sigma}^T\,\boldsymbol{\vartheta}=\tilde{\mathbf{F}}^i_\xi\,-\tilde{\mathbf{F}}^i_\eta.
\end{equation}
Here $\tilde{\mathbf{F}}^i$ is the tangent map projected on the interface. It is important to note that Eqs.~\eqref{eq:burgers_interface_reference} and \eqref{eq:burgers_interface_current} represent the damage contents due to delamination measured from the reference and the current configuration of the body. Thus, these measures are dependent on the choice of the configuration. In the context of measuring GND density, Cermelli and Gurtin~(2001)~\cite{cermelli2001characterization} argued that the measures of damage content (dislocation density in the original paper) must be invariant ``under superposed compatible elastic deformations" as well as ``compatible local changes in reference configuration", as discussed in \S~\ref{sec:Continuous and discontinuous manifold}. Therefore, we also need to provide a suitable damage density due to delamination which satisfies these conditions. Gupta and Steigmann~(2012)~\cite{gupta2012plastic} also provided a similar measure for the interface dislocation density, which they referred to as the \emph{true interface dislocation density}. 

To provide a measure of damage density that remains invariant under superposed local compatible changes in the reference configuration~\cite{cermelli2001characterization}, let us first consider two reference configurations differed by a smooth, local, compatible, tangent map $\mathbf{H}$ such that $d\mathbf{X}_2=\mathbf{H}\,d\mathbf{X}_1$. Here $d\mathbf{X}_1$ and $d\mathbf{X}_2$ denote the infinitesimal fiber in the two reference configurations, respectively. Under compatible deformation, the total damage content measured from both the reference configurations must be the same. Therefore, one can write
\begin{equation}\label{eq:equal_ref}
  \int_{\Omega_1}(\text{Curl}_1\,\mathbf{F}^i_1)^T\,\mathbf{N}\,dA_1-\int_{\Gamma_1}\llbracket\mathbf{F}^i_1\rrbracket d\mathbf{X}_1=\int_{\Omega_2}(\text{Curl}_2\,\mathbf{F}^i_2)^T\,\mathbf{N}\,dA_2-\int_{\Gamma_2}\llbracket\mathbf{F}^i_2\rrbracket d\mathbf{X}_2.
\end{equation}
A routine calculation (see~\ref{sec:appendix_del_total}) shows that the true measure of the damage density corresponding to delamination $\tilde{\boldsymbol{\Sigma}}_\delta$, projected on the interface, can be written as
\begin{equation}\label{eq:true_interface_dislocation}    \tilde{\mathbf{\Sigma}}_\delta=\mathbf{\Sigma}^T\boldsymbol{\vartheta}\,\tilde{\mathbf{F}}^{i^{-1}}_\delta=\boldsymbol{\sigma}^T\boldsymbol{\vartheta}\,\tilde{\mathbf{F}}^{e}_\delta,\quad\delta=\xi,\eta.
\end{equation}
Although these measures are not exactly the same, we show that they are related through a second-order tensor similar to the relative distortion tensor in~\S~\ref{sec:debonding}. To demonstrate this, we rewrite Eq.~\eqref{eq:true_interface_dislocation} as 
\begin{equation}\label{eq:true_interface_dislocation_2}
\tilde{\mathbf{\Sigma}}_\xi\tilde{\mathbf{F}}^i_\xi\tilde{\mathbf{F}}^{i^{-1}}_\eta=\mathbf{\Sigma}^T\boldsymbol{\vartheta}\tilde{\mathbf{F}}^{i^{-1}}_\eta.
\end{equation}
Now let us introduce the relative distortion tensor at the interface between the two lamin$\ae$ as 
\begin{equation}\label{eq:relative_distorsion_tensor_Interface_del}
     {\tilde{\mathbf{M}}} = {\tilde{\mathbf{F}}^{i}_\xi}\,\tilde{\mathbf{F}}^{i^{-1}}_\eta.
   \end{equation}
Now substituting the definition of ${\tilde{\mathbf{M}}}$ into Eq.~\eqref{eq:true_interface_dislocation_2}, it can be shown that the damage densities due to delamination measured from the two lamin$\ae$ are related through 
\begin{equation}\label{eq:relative_distorsion_tensor_Interface_del_1}
    \tilde{\mathbf{\Sigma}}_\xi\,\tilde{\mathbf{M}}=\tilde{\mathbf{\Sigma}}_\eta.
\end{equation}
Note that although the idea of a relative distortion tensor in Eq.~\eqref{eq:relative_distorsion_tensor_Interface} and ~\eqref{eq:relative_distorsion_tensor_Interface_del} are similar, they carry different physical quantities. 

\section{Geometric interpretation of the measures of damage}\label{sec:Geometric interpretation of the measures of damage}

In \S~\ref{Damage characterization}, the measures of damage have been obtained using only physical arguments. In this section, we provide a geometric interpretation of these measures. As mentioned earlier in ~\S~\ref{sec:Kinematics}, a body can be considered as a differentiable manifold. Therefore, one can also characterize the defects from a more geometric perspective by studying the local metrics, connections or curvatures in the material manifold. Our approach closely follows the work of Clayton (2012)~\cite{clayton2012anholonomic} and Paul and Freed (2020)~\cite{paul2020characterizing}.

\subsection{Matrix cracking and fiber breakage}
We first start with the geometric interpretation of matrix cracking. Let us define a set of base vectors $\mathbf{E}^A$ that span the reference configuration of the body. The tangent map $\mathbf{F}^d_m$ induces a natural base vector $\mathbf{e}_{m_{a}}$ in the configuration $\kappa^d_m(\mathcal{B}_0)$. Thus, the tangent map $\mathbf{F}^d_m$ can be written in the coordinate frame as
\begin{equation}\label{eq:tangent_map_notation}
    \mathbf{F}^d_m={F}^{d^{~a}}_{m_{~A}}\,(\mathbf{X},t)\,\mathbf{e}_{m_{a}}\otimes\mathbf{E}^A.
\end{equation}
The convected base vectors and their reciprocals in the reference configuration can be defined as 
\begin{equation}\label{eq:convected_base_vector1}
   \overline{\mathbf{E}}^A(\mathbf{x}^d_m,t)={F}^{d^{-1^{~A}}}_{m_{~a}}\,(\mathbf{x}^d_m,t)\,\mathbf{e}^a_{m},\quad
   \overline{\mathbf{E}}_A(\mathbf{X},t)={F}^{d^{~a}}_{m_{~A}}\,(\mathbf{X},t)\,\mathbf{e}_{m_{a}}\quad\text{such that}\quad \overline{\mathbf{E}}^A\cdot\overline{\mathbf{E}}_B=\delta^A_B
\end{equation}
where $\delta^A_B$ represents the Kr\"onecker delta. With the help of the convected base vectors in Eq.~\eqref{eq:convected_base_vector1}, we determine the metric corresponding to the tangent map $\mathbf{F}^d_m$ in the reference configuration $\kappa_R(\mathcal{B}_0)$ as 
\begin{equation}
    {C}^d_{m_{AB}}\defeq\mathbf{\overline{E}}_A\,\cdot \mathbf{\overline{E}}_B= {F}^{d^{\,a}}_{m_{\,A}}\,{F}^{d^{\,a}}_{m_{\,B}}.
\end{equation}
Now we introduce an appropriate linear connection with respect to the metric $\mathbf{C}^d_m$ along with its covariant derivative. In the reference configuration, the covariant derivative of a vector field $\mathbf{W}$ in the direction of another vector field $\mathbf{V}$ is given as~\cite{marsden1994mathematical} 
\begin{equation}    
\nabla_{\mathbf{V}} \mathbf{W} 
= \left( V^B \partial_B W^A + \Gamma^A_{BC} W^C V^B \right)\mathbf{E}_A .
\end{equation}
Let $\mathbf{\Gamma}$ denote the connection coefficient (Christoffel symbol) of the reference configuration $\kappa_R(\mathcal{B}_0)$, which satisfies the identity 
\begin{equation}\label{eq:connection_coefficient_gradient}
    \Gamma^A_{BC} \partial_A = \nabla_{\partial_B} \partial_C.
\end{equation}
Using Eq.~\eqref{eq:connection_coefficient_gradient}, one can write
\begin{equation}
    \partial_B \overline{\mathbf{E}}_{A}= \partial_B {F}^{d^{\,a}}_{m_{\,A}} \, \mathbf{e}_{m_{\,a}} = {F}^{d^{{-1}^{\,D}}}_{m_{\,a}} \, \partial_B {F}^{d^{\,a}}_{m_{\,A}} \, \overline{\mathbf{E}}_D.
\end{equation}
If $\overset{\mathbf{C}^d_m}{\mathbf{\Gamma}}$ denotes the connection coefficient associated with the metric $\mathbf{C}^d_m$, using Eq.~\eqref{eq:convected_base_vector1} along with the identity of  $\partial_B\overline{\mathbf{E}}_A=\overset{\mathbf{C}^d_m}{\Gamma}{{^D_{B\,A}}}\overline{\mathbf{E}}_D$, we obtain 
\begin{equation}
    \overset{\mathbf{C}^d_m}{\Gamma}{^D_{B\,A}}={F}^{d^{{-1}^D}}_{m_{\,a}}\partial_B{F}^{d^{\,a}}_{m_{\,A}}.
\end{equation}
Since $\mathbf{F}^d_m$ is incompatible, the connection coefficient need not be symmetric. The skew-symmetric part of the connection coefficient represents the torsion associated with this connection which is given as
\begin{equation}\label{eq:torsion_m}
    T_m{^{D}_{AB}} = \frac{1}{2} \left(\overset{\mathbf{C}^d_m}{\Gamma}{^D_{B\,A}}-\overset{\mathbf{C}^d_m}{\Gamma}{^D_{A\,B}}\right)
= \frac{1}{2} \, {F}^{d^{{-1}^{\, D}}}_{m_{\,a}}\left( \partial_B {F}^{d^{\, a}}_{m_{\,A}}\,- \partial_A{F}^{d^{\,a}}_{m_{\,B}}\right)\neq 0.
\end{equation}
The torsion $\mathbf{T}_m$ in Eq.~\eqref{eq:torsion_m} is the geometric measure of incompatibility in the matrix configuration $\mathbf{\kappa}^d_m(\mathcal{B}_0)$. A routine calculation {(see \ref{sec:Appendix_matrix_crack} for a detailed calculation)} shows that the components of the torsion tensor have a one-to-one correspondence with the components of the matrix crack density tensor $\mathbf{G}_m$.

Similarly, a similar geometric interpretation can be obtained for the fiber breakage by following the same procedure for the configuration $\kappa^d_f(\mathcal{B}_0)$. In this case, the torsion of the connection in the fiber configuration $\mathbf{\kappa}^d_f(\mathcal{B}_0)$ can be derived as 
\begin{equation}\label{eq:torsion}
    T_f{^{D}_{AB}} = \frac{1}{2} \left(\overset{\mathbf{C}^d_f}{\Gamma}{^D_{B\,A}}-\overset{\mathbf{C}^d_f}{\Gamma}{^D_{A\,B}}\right)
= \frac{1}{2} \, {F}^{d^{{-1}^{\, D}}}_{f_{\,p}}\left(\partial_B {F}^{d^{\,p}}_{f_{\,A}}\,- \partial_A{F}^{d^{\,p}}_{f_{\,B}}\right).
\end{equation}
Following a similar argument, the torsion tensor $\mathbf{T}_f$ provides the necessary geometric interpretation for the fiber breakage density tensor $\mathbf{G}_f$. 

{\subsection{Debonding and interfacial slip}\label{sec:geo_debonding}

For the geometric interpretation of debonding and interfacial slip, let us first consider a small neighborhood $\mathcal{B}_p$ around a point $p$ on the matrix-fiber interface in the reference configuration of the body such that $\mathcal{B}_p=\mathcal{B}_{p_m}\cup S_p \cup\mathcal{B}_{p_f}$. $\mathcal{B}_{p_m}$ and $\mathcal{B}_{p_f}$ are subdomains of the neighborhood containing the matrix and the fiber phases respectively, that intersect at an interface ${S}_p$. Following our discussions in~\S~\ref{sec:debonding}, we consider \emph{locally homogeneous} motions $\boldsymbol{\zeta}_{\alpha},\: \alpha=m, f$ on $\mathcal{B}_{p_{\alpha}}$ such that $\mathbf{F}^d_{\alpha}=T\boldsymbol{\zeta}_{\alpha}$. The spatial velocities for each constituent phase, $\mathbf{v}^d_\alpha$ can be defined as vector fields in the local tangent spaces of $\boldsymbol{\zeta}_{\alpha}(\mathcal{B}_{p_{\alpha}})$ at point $p$ such that $\mathbf{v}^d_{\alpha}: \boldsymbol{\zeta}_{\alpha}(\mathcal{B}_{p_{\alpha}})\rightarrow \mathbb{R}^3$~\cite{marsden1994mathematical} and, $\mathbf{v}^d_{\alpha}=\partial\boldsymbol{\zeta}_{\alpha}/\partial t\in T\boldsymbol{\zeta}_{\alpha}(\mathcal{B}_{p_{\alpha}})$. A pull-back operation on the spatial velocities leads to
\begin{equation}\label{eq:velocities_pull_back}
\mathbf{V}_{\alpha}=\boldsymbol{\zeta}_{\alpha}^*\mathbf{v}_{\alpha}=\mathbf{F}^{d^{-1}}_{\alpha}\,\mathbf{v}_{\alpha}.
\end{equation}
$\mathbf{V}_{\alpha}$ are the material descriptions of the respective spatial velocities and thus, $\mathbf{V}_{\alpha}\in T\mathcal{B}_p$. The jump in the pulled-back spatial velocities across the matrix-fiber interface $S_p$ can be defined as $\llbracket\mathbf{V}\rrbracket\defeq \mathbf{V}_m-\mathbf{V}_f$. To understand the meaning of this jump, let us write the vector field $\mathbf{V}\in T\mathcal{B}_p$ as
\begin{equation}\label{eq:V_distributional}
    \mathbf{V}=h_m\,\mathbf{V}_m\,+\,h_f\,\mathbf{V}_f\quad\text{with}\quad h_{\alpha}\cdot h_{\alpha}=h_{\alpha}, h_m\cdot h_f=0\:\text{and,}\: h_m+h_f=1.
\end{equation}
Here $h_{\alpha}$ are the Heaviside step functions on both sides of the interface $S_p$. Now the gradient of the vector field $\mathbf{V}$ can be written as
\begin{equation}\label{eq:grad_V}
\begin{aligned}
    \nabla_X\mathbf{V}\big|^i_J&=h_m\,\partial_J V^{\,i}_m+h_f\,\partial_J V^{\,i}_f+(\partial_J h_m)\,V^{\,i}_m+(\partial_J h_f)\,V^{\,i}_f\\&=\underbrace{h_m\nabla_X\mathbf{V}_m\big|^i_J\,+\,h_f\nabla_X\mathbf{V}_f\big|^i_J}_{\text{bulk}}\,+\,\underbrace{(V^i_m-V^i_f)}_{\llbracket\mathbf{V}\rrbracket^i}\,{N}_J\,\Delta_{S_p}.
\end{aligned}
\end{equation}
The last of Eq.~\eqref{eq:grad_V} follows from the derivatives of the Heaviside step functions $\partial_Jh_m=-\partial_Jh_f={N}_J\Delta_{S_p}$ where $\Delta_{S_p}$ and ${\mathbf{N}}$ are the Dirac distribution at the interface $S_p$ and the unit normal to the interface at a point $\mathbf{X}\in S_p$, respectively. Since we are interested in determining the interfacial slip rate, the pulled-back velocities and their jump need to be projected on the interface. For this purpose, let us define a second-order projection tensor $\bar{\mathbf{1}}:T_X\mathcal{B}_p\rightarrow T_XS_p$ as
\begin{equation}\label{eq:projection_interface}
    \bar{\mathbf{1}}=\mathbf{I}-\mathbf{N}\otimes\mathbf{N}.
\end{equation}
Using the projection tensor $\bar{\mathbf{1}}$ on the jump in the velocity field $\llbracket\mathbf{V}\rrbracket$, the interfacial slip rate $\boldsymbol{\Lambda}\in T_{X}S_p$ is obtained as
\begin{equation}\label{eq:rel_vel_interface}
    \boldsymbol{\Lambda}\defeq \bar{\mathbf{1}}\llbracket\mathbf{V}\rrbracket= \bar{\mathbf{1}}(\mathbf{V}_m-\mathbf{V}_f)\quad \text{with}\quad \bar{\mathbf{1}}\mathbf{V}_{\alpha}=\mathbf{V}_{\alpha}-(\mathbf{N}\otimes\mathbf{N})\mathbf{V}_{\alpha}=\mathbf{V}_{\alpha}-(\mathbf{V}_{\alpha}\cdot\mathbf{N})\mathbf{N}.
\end{equation}
From the last of Eq.~\eqref{eq:rel_vel_interface}, one can easily notice that when the pulled-back velocity $\mathbf{V}_{\alpha}$ is projected onto the interface using $\bar{\mathbf{1}}$, the resulting velocity is the tangential component of $\mathbf{V}_{\alpha}$ along the matrix-fiber interface. Thus, the interfacial slip rate, defined in Eq.~\eqref{eq:interfacial_slip_rate}, can be interpreted as the jump in the vector field $\mathbf{V}$ projected onto the interface.}

\subsection{Delamination}

To characterize delamination, we consider an interface $S_R$ in the body which separates the two lamin\ae~denoted by $\xi$ and $\eta$ as shown in \S~\ref{sec:delamination}. Due to the delamination, there is a jump in the displacement field across the interface $S_R$. To accommodate the jump in the displacement field, the tensor field $\mathbf{F}^i$ can be written as~\cite{vignolo2018junction} 
\begin{equation}\label{eq:jump_F_i}
\mathbf{F}^i=h_\xi\,\mathbf{F}^{i}_\xi+h_\eta\,\mathbf{F}^{i}_\eta
\end{equation}
where $h_{\delta}$, $\delta=\xi\;\text{or}\;\eta$ are the Heaviside step functions on both sides of the interface $S_R$ with the properties 
\begin{equation}\label{eq:Lambda_prop}
    h_{\delta}\cdot h_{\delta}=h_{\delta},\quad h_{\xi}\cdot h_{\eta}=h_{\eta}\cdot h_{\xi}=0,\quad \text{and}\quad h_{\xi}+h_{\eta}=1.
\end{equation} 
A differentiation of Eq.~\eqref{eq:jump_F_i} with respect to the reference coordinate leads to
\begin{equation}\label{eq:product_rule}
    \partial_B{F}^{i^{\,a}}_{\,A}=\left(\partial_B h_\xi\right)\,{F}^{i^{\,a}}_{\xi_{\,A}}+h_\xi\,\partial_B{F}^{i^{\,a}}_{\xi_{\,A}}+(\partial_Bh_\eta)\,{F}^{i^{\,a}}_{\eta_{\,A}}+h_\eta\,\partial_B{F}^{i^{\,a}}_{\eta_{\,A}}.
\end{equation}
Through a routine calculation (see~\ref{sec:appendixC}), $\partial_B{F}^{i^{\,a}}_{\,A}$ can be written as
\begin{equation}\label{eq:convected_derivative_F_i}  \partial_B{F}^{i^{\,a}}_{\,A}=h_\xi\,\partial_B{F}^{i^{\,a}}_{\xi_{\,A}}+h_\eta\,\partial_B{F}^{i^{\,a}}_{\eta_{\,A}}+n_B\Delta_{S_R}\,\llbracket{F}^i\rrbracket^{\,a}_{\,A}.
\end{equation}
Here $\Delta_{S_R}$ is the Dirac delta distribution at the interface $S_R$ and $\mathbf{n}$ is the unit normal to the interface $S_R$. It can be observed that the last term in the right-hand side of Eq.~\eqref{eq:convected_derivative_F_i} is the extra singular term proportional to the jump across the interface.

Now let $\overset{\mathbf{C}^i}{\mathbf{\Gamma}}$ denote the connection coefficient associated with the metric $\mathbf{C}^i$. With the help of the convected base vectors (similar to Eq.~\eqref{eq:convected_base_vector1}) and using the identity $\partial_B\overline{{E}}_A=\overset{\mathbf{C}^i}{\Gamma}{{^D_{B\,A}}}\overline{{E}}_D$, we define the connection coefficient as 
\begin{equation}\label{eq:connection_coff_natural}
    \overset{\mathbf{C}^i}{\Gamma}{^D_{B\,A}}={F}^{i^{{-1}^D}}_{\,a}\partial_B{F}^{i^{\,a}}_{\,A}.
\end{equation}
Substituting the expressions of $\mathbf{F}^i$ and its derivative with respect to the reference coordinates from Eqs.~\eqref{eq:jump_F_i} and \eqref{eq:convected_derivative_F_i} respectively into Eq.~\eqref{eq:connection_coff_natural}, the connection coefficient corresponding to the metric $\mathbf{C}^i$ is evaluated as 
\begin{equation}\label{eq:metric_C^i}
    \overset{\mathbf{C}^i}{\Gamma}{^D_{B\,A}}=(h_\xi\,{F}^{i^{-1}}_\xi+h_\eta\,{F}^{i^{-1}}_\eta)^D_a\,(h_\xi\,\partial_B{F}^{i^{\,a}}_{\xi_{\,A}}+h_\eta\partial_B{F}^{i^{\,a}}_{\eta_{\,A}}+n_B\Delta_{S_R}\,\llbracket{F}^i\rrbracket^{\,a}_{\,A}).
\end{equation}
Through a routine calculation using the properties~\eqref{eq:Lambda_prop}, the connection co-efficient can be written as
\begin{equation}\label{eq:jump_connection}
    \overset{\mathbf{C}^i}{\Gamma}{^D_{\,BA}}=h_\xi\, \overset{\mathbf{C}^i}{\Gamma}_\xi{^D_{\,BA}}+h_\eta\, \overset{\mathbf{C}^i}{\Gamma}_\eta{^D_{\,BA}}+({F}^{i^{-1}})^D_{a}\,n_B\Delta_{S_R}\llbracket{F}^{i}\rrbracket^{\,a}_A.
\end{equation}
The torsion associated with the connection coefficient is finally evaluated as
\begin{equation}\label{eq:jump_connection1}
\begin{aligned}
     T^D_{AB}&=\underbrace{h_\xi\,\bigg( \underbrace{\frac{1}{2} \, ({F}^{i^{{-1}}}_{\xi})^{\, D}_{\,a}\left( \partial_B {F}^{i^{\,a}}_{\xi_{\,A}}\,- \partial_A{F}^{i^{\,a}}_{\xi_{\,B}}\right)}_{{\mathbf{T}_{\xi}}}\bigg)+h_\eta\,\bigg(\underbrace{\frac{1}{2} \, ({F}^{i^{{-1}}}_{\eta})^{\,D}_{\,a}\left(\partial_B {F}^{i^{\,a}}_{\eta_{\,A}}\,- \partial_A{F}^{i^{\,a}}_{\eta_{\,B}}\right)}_{{\mathbf{T}_{\eta}}}\bigg)}_{{Bulk}}\\&+\underbrace{\underbrace{{\frac{1}{2}({F}^{i^{-1}})^D_a\,\Delta_{S_R}\Bigg(\,\,n_B\,\llbracket{F}^{i}\rrbracket\,^a_A-\,n_A\,\llbracket{F}^{i}\rrbracket\,^a_B\,\Bigg)}}_{\mathbf{T}\big|_{S_R}}}_{\text{Interface}}.
     \end{aligned}
\end{equation}
The first and second terms in the right-hand side of Eq.~\eqref{eq:jump_connection1} provide the torsion corresponding to the bulk parts of the body (i.e., excluding the interface), whereas the last term is the required measure of incompatibility (torsion) corresponding to the jump in the displacement field across the interface $S_R$. Therefore, the total torsion can  be written as
\begin{equation}\label{eq:torsion_del}
    T^D_{AB}=h_\xi\,{T_{\xi}}^{D}_{AB}+h_\eta\,{T_{\eta}}^{D}_{AB}+T^D_{AB}\bigg|_{S_R}
\end{equation}
where the measure of incompatibility corresponding to the interface is given as
\begin{equation}\label{eq:tosion_int_del}
    T^D_{AB}\Bigg|_{S_R}=\frac{1}{2}({F}^{i^{-1}})^D_a\,\Delta_{S_R}\Bigg(\,\,n_B\,\llbracket{F}^{i}\rrbracket\,^a_A-\,n_A\,\llbracket{F}^{i}\rrbracket\,^a_B\,\Bigg).
\end{equation}
The torsion $T^D_{AB}$ in Eq.~\eqref{eq:jump_connection1} is the required geometric measure of incompatibility in the natural configuration $\kappa_i(\mathcal{B}_0)$. It can be shown {(c.f.,~\ref{sec:Appendix_delamination})} that the interface part of the torsion tensor have a one-to-one correspondence with its corresponding part of the total damage density accounting for delamination $\mathbf{b}_R$ in Eq.~\eqref{eq:burgers_interface_reference}.

{\section{Discussions}\label{sec:discussion}

\subsection{Incorporation of the damage measures into a constitutive framework}\label{sec:constitutive}

In this section, we develop a general constitutive structure based on the presented kinematic framework. Since our kinematic framework is derived within the theory of multiple natural configurations, we follow the constitutive modeling strategy developed by Rajagopal and Srinivasa~(2004)~\cite{rajagopal2004thermomechanical}. We assume the existence of a Helmholtz free energy, $\psi$ that describes the elastic response of the body for a given natural configuration, $\kappa_i(\mathcal{B}_0)$. Therefore, $\psi$ can be assumed as a function of the elastic deformation gradient $\mathbf{F}^e$ or, alternatively $\mathbf{F}$ and $\mathbf{F}^i$. Moreover, the measures of the microstructural defects in the form of matrix cracks, fiber breakage, interfacial slip and delamination are crucial for the descriptions of the natural configuration from which the elastic response is measured and hence, these measures enter into the functional form of $\psi$ as internal state variables. In view of the material frame indifference, the functional form for the Helmholtz free energy for a Lagrangian formulation can be written as
\begin{equation}\label{eq:helmholtz_energy}
    \psi=\hat{\psi}(\mathbf{F}^e,\overline{\mathbf{G}}{_{\alpha_\delta}},\overline{\boldsymbol{\Lambda}}_{\delta},{\hat{\mathbf{\Sigma}}})= \bar{\psi}\,(\mathbf{C}, \mathbf{C}^i,\overline{\mathbf{G}}{_{\alpha_\delta}},\overline{\boldsymbol{\Lambda}}_{\delta},{\hat{\mathbf{\Sigma}}}),\qquad~\alpha=m, f\:\text{and,}\:\delta=\xi,\eta.
\end{equation}
Since the damage measures $\mathbf{G}_{\alpha}$ are defined in the natural configuration in our theory, these mixed second-order tensors are pulled back into the reference configuration $\kappa_R(\mathcal{B}_0)$ via $\overline{\mathbf{G}}_{\alpha}=\mathbf{F}^{d^{-1}}_{\alpha}\,\mathbf{G}_{\alpha}\,\mathbf{F}^{d}_{\alpha}$. $\overline{\boldsymbol{\Lambda}}_{\delta}$ represents the amount of interfacial slip at time $t$ in the reference configuration $\kappa_R(\mathcal{B}_0)$ such that $\overline{\boldsymbol{\Lambda}}_{\delta}= \int_0^t{\boldsymbol{\Lambda}}_{\delta}\,\mathrm{d}t)$. We assume that the elastic response of the material is that of a Green elastic solid. Therefore, the second Piola-Kirchhoff stress can be written as    
\begin{equation}\label{eq:green_elastic_solid}
     \mathbf{S}=2\,\rho_0\,\dfrac{\partial\bar{\psi}}{\partial\mathbf{C}}.
\end{equation}
where $\rho_0$ is the material density in the reference configuration of the body. If $\Pi$ denotes the rate of dissipation, from the balance of energy, one can write 
\begin{equation}\label{eq:thermodynamic_equation}
    \Pi=\dfrac{1}{2}\,\mathbf{S}\boldsymbol{:}\dot{\mathbf{C}}-\rho_0\,\dot{\psi} \ge 0.
\end{equation}

In a traditional Coleman-Noll procedure~\cite{noll1967materially}, the evolution equations for the damage measures are obtained in such a way that some form of the second law of thermodynamics such as the Clausius-Duhem inequality is satisfied. Rajagopal and Srinivasa~(2004)~\cite{rajagopal2004thermomechanical} proposed a different approach in which a kinematic variable is considered to be admissible when it ensures that the rate of dissipation is non-negative. Thereafter, the evolution equation for the kinematic variable is obtained through a maximization of the rate of dissipation function, subjected to the constraint~\eqref{eq:thermodynamic_equation}. Therefore, the functional form of the rate of dissipation function needs to be assumed first in this procedure. Since the rate of dissipation $\Pi$ governs the evolution of the natural configuration $\kappa_i(\mathcal{B}_0)$, we assume its functional form as
\begin{equation}\label{eq:function_dissipation}
    \Pi=\overline{\Pi}\,(\mathbf{C}^i, \dot{\mathbf{C}}^i,\overline{\mathbf{G}}{_{\alpha_\delta}},\dot{\overline{\mathbf{G}}}{_{\alpha_\delta}},\overline{\boldsymbol{\Lambda}}_{\delta},\boldsymbol{\Lambda}_{\delta},\hat{\mathbf{\Sigma}},\dot{\hat{\mathbf{\Sigma}}}).
\end{equation}
Now with the help of the Eqs.~\eqref{eq:helmholtz_energy}, \eqref{eq:green_elastic_solid} and \eqref{eq:thermodynamic_equation}, one can write
\begin{equation}\label{eq:constrain}
    \Pi=-\rho_0\Bigg[\dfrac{\partial \bar{\psi}}{\partial \mathbf{C}^i}\boldsymbol{:}\dot{\mathbf{C}}^i+ \dfrac{\partial \bar{\psi}}{\partial \overline{\mathbf{G}}_{\alpha_\delta}}\boldsymbol{:}\dot{\overline{\mathbf{G}}}_{\alpha_\delta}+\dfrac{\partial \bar{\psi}}{\partial \overline{{\boldsymbol{\Lambda}}}_{\delta}}\boldsymbol{\cdot}{{{\boldsymbol{\Lambda}}}}_{\delta}+\dfrac{\partial \bar{\psi}}{\partial \hat{\mathbf{\Sigma}}}\boldsymbol{:}\dot{\hat{\mathbf{\Sigma}}}\Bigg].
\end{equation}
To obtain the evolution equations, the rate of dissipation $\Pi$ is maximized, subject to the constraint~\eqref{eq:constrain}, using a Lagrange multiplier technique. The Lagrangian for this constrained maximization is written as
\begin{equation}
    \mathcal{L}={\Pi}\,+\overline{m}\Bigg(\,\Pi+\rho_0\Bigg[\dfrac{\partial \bar{\psi}}{\partial \mathbf{C}^i}\boldsymbol{:}\dot{\mathbf{C}}^i+\dfrac{\partial \bar{\psi}}{\partial \overline{\mathbf{G}}_{\alpha_\delta}}\boldsymbol{:}\dot{\overline{\mathbf{G}}}_{\alpha_\delta}+\dfrac{\partial \bar{\psi}}{\partial\overline{\boldsymbol{\Lambda}}_{\delta}}\boldsymbol{\cdot}{{\boldsymbol{\Lambda}}}_{\delta}+\dfrac{\partial \bar{\psi}}{\partial \hat{\mathbf{\Sigma}}}\boldsymbol{:}\dot{\hat{\mathbf{\Sigma}}}\Bigg]\Bigg).
\end{equation}
where $\bar{m}$ is the Lagrange multiplier that can be obtained from the satisfaction of the constraint equation~\eqref{eq:constrain} as $\bar{m}/(1+\bar{m})=1$. Now, carrying out the optimization, we obtain the set of evolution equations as
\begin{align}
    &\frac{\partial}{\partial\dot{\mathbf{C}}^i}\Bigg[\Pi+\Bigg(\rho_0\,\dfrac{\partial\overline{\psi}}{\partial\overline{\mathbf{G}}_{\alpha_\delta}}\,\boldsymbol{:}\dot{\overline{\mathbf{G}}}_{\alpha_\delta}\Bigg)+\Bigg(\rho_0\,\dfrac{\partial\overline{\psi}}{\partial\overline{\boldsymbol{\Lambda}}_{\delta}}\,\boldsymbol{\cdot}{{\boldsymbol{\Lambda}}}_{\delta}\Bigg)+\Bigg(\rho_0\,\frac{\partial\overline{\psi}}{\partial\hat{\mathbf{\Sigma}}}\,\boldsymbol{:}\dot{\hat{\mathbf{\Sigma}}}\Bigg)\Bigg]=-\rho_0\,\frac{\partial\overline{\psi}}{\partial\mathbf{C}^i}\label{eq:evolution1}\\
    &\frac{\partial}{\partial\dot{\overline{\mathbf{G}}}_{\alpha_\delta}}\Bigg[\Pi+\Bigg(\rho_0\,\frac{\partial\overline{\psi}}{\partial\mathbf{C}^i}\boldsymbol{:}\dot{\mathbf{C}}^i\Bigg)+\Bigg(\rho_0\,\frac{\partial\overline{\psi}}{\partial\overline{\boldsymbol{\Lambda}}_{\delta}}\,\boldsymbol{\cdot}{{\boldsymbol{\Lambda}}}_{\delta}\Bigg)+\Bigg(\rho_0\,\frac{\partial\overline{\psi}}{\partial\hat{\mathbf{\Sigma}}}\,\boldsymbol{:}\dot{\hat{\mathbf{\Sigma}}}\Bigg)\Bigg]=-\rho_0\,\frac{\partial\overline{\psi}}{\partial\overline{\mathbf{G}}_{\alpha_\delta}}\label{eq:evolution2}\\
     &\frac{\partial}{\partial{{\boldsymbol{\Lambda}}}_{\delta}}\Bigg[\Pi+\Bigg(\rho_0\,\frac{\partial\overline{\psi}}{\partial\mathbf{C}^i}\boldsymbol{:}\dot{\mathbf{C}}^i\Bigg)+\Bigg(\rho_0\,\frac{\partial\overline{\psi}}{\partial\overline{\mathbf{G}}_{\alpha_\delta}}\,\boldsymbol{:}\dot{\overline{\mathbf{G}}}_{\alpha_\delta}\Bigg)+\Bigg(\rho_0\,\frac{\partial\overline{\psi}}{\partial\hat{\mathbf{\Sigma}}}\,\boldsymbol{:}\dot{\hat{\mathbf{\Sigma}}}\Bigg)\Bigg]=-\rho_0\,\frac{\partial\overline{\psi}}{\partial\overline{\boldsymbol{\Lambda}}_{\delta}}\label{eq:evolution3}\\
     &\frac{\partial}{\partial\dot{\hat{\mathbf{\Sigma}}}}\Bigg[\Pi+\Bigg(\rho_0\,\frac{\partial\overline{\psi}}{\partial\mathbf{C}^i}\boldsymbol{:}\dot{\mathbf{C}}^i\Bigg)+\Bigg(\rho_0\,\frac{\partial\overline{\psi}}{\partial\overline{\mathbf{G}}_{\alpha_\delta}}\,\boldsymbol{:}\dot{\overline{\mathbf{G}}}_{\alpha_\delta}\Bigg)+\Bigg(\rho_0\,\frac{\partial\overline{\psi}}{\partial\overline{\boldsymbol{\Lambda}}_{\delta}}\,\boldsymbol{\cdot}{{\boldsymbol{\Lambda}}}_{\delta}\Bigg)\Bigg]=- \rho_0\,\frac{\partial\overline{\psi}}{\partial\hat{\mathbf{\Sigma}}}.\label{eq:evolution4}
\end{align}
The set of Eqs.~\eqref{eq:evolution2}--\eqref{eq:evolution4} are the required evolution equations for the derived damage measures. 

Let us now understand the physical meanings of the obtained evolution equations. Using Eshelby's procedure~\cite{eshelby1951force} of determining force on an inhomogeneity, it can be shown that the terms $-\rho_0\partial\psi/\partial\mathbf{C}^i$, $-\rho_0\partial\psi/\partial\mathbf{G}_{\alpha_{\delta}}$, $-\rho_0\partial\psi/\partial\overline{\boldsymbol{\Lambda}}_{{\delta}}$ and $-\rho_0\partial\psi/\partial\hat{\mathbf{\Sigma}}$ represent the configurational forces (Eshelby energy-momentum tensor) that are thermodynamic conjugates to the corresponding damage measures (c.f.,~\cite{rajagopal2005role,paul2022use} for a detailed derivation). Hence, the scalar product between the configurational forces and the rates of the damage measures yield the rate of dissipation associated with that damage mechanism. For example, $-\rho_0\,{\partial\overline{\psi}}/{\partial\hat{\mathbf{\Sigma}}}\,\boldsymbol{:}\dot{\hat{\mathbf{\Sigma}}}$ represents the rate of dissipation associated with delamination. In view of the physical meanings of these terms, one can observe that the left hand sides of the evolution equations~\eqref{eq:evolution1}--\eqref{eq:evolution4} are derivatives of the rate of dissipation \emph{available for the specific dissipative process} with respect to the rate of the corresponding kinematic variable. Therefore, the evolution equations are implicit equations in the respective damage measures.

The physical meanings of the evolution equations have an important consequence. It is worth noting that although different damage mechanisms are treated separately in our developed kinematic framework, their evolution equations are a set of coupled equations. The coupling between the evolution equations allows us to model the interaction between different damage mechanisms, such as matrix crack-induced delamination. Such phenomena can be modeled by simply choosing a suitable form for the Helmholtz free energy and the rate of dissipation function. 

\subsection{Relations with existing damage measures and models}

In this section, we compare the derived damage variables with some of the existing damage measures and modeling techniques. For this comparison, we consider a linearized theory such that $\underset{\forall \mathbf{X}}{\text{max}}\norm{\partial\mathbf{
u/\partial\mathbf{X}}}=\mathcal{O}(\upsilon),\,\upsilon\ll 1$ and homogeneous deformations.

To characterize matrix cracking, Talreja~(1985)~\cite{talreja1985continuum} used the crack density tensor of Kachanov~(1980)~\cite{kachanov1980continuum} with necessary interpretations. In this work, a vector $\mathbf{p}^g=d^g\,\mathbf{n}^g$ was assigned to each crack in a representative volume element with volume $\mathcal{V}$ such that $d=f(a^g, dA^g)$ where $dA^g$ and $a^g$ represent the crack area and its characteristic dimension, respectively. From the discussion in~\S~\ref{sec:Continuous and discontinuous manifold}, it is evident that the vector $\mathbf{p}^g$ is the jump in the displacement field across the crack surface. Hence, following the derivation in~\ref{sec:Appendix_matrix_crack}, one can write 
\begin{equation}\label{eq:crack_relation}
     \mathbf{p}^g=d^g\,\mathbf{n}^g=\text{Curl}\,\mathbf{F}^d_m\, d\mathbf{a}_m.
\end{equation}
Now using the definition of the matrix crack density tensor $\mathbf{G}_m$ in Eq.~\eqref{eq:matrix_crack_density}, we obtain
\begin{equation}\label{eq:talreja_crack2}    
\mathbf{p}^g=(J^d_m\,\mathbf{F}^{d^{-1}}_m\,\mathbf{G}_m\,d\mathbf{a}_m)^g\simeq(J^d_m\,\mathbf{G}_m\,d\mathbf{a}_m)^g.
\end{equation}
In the last of Eq.~\eqref{eq:talreja_crack2}, we have used the assumption of a small displacement gradient such that terms of order $\mathcal{O}(\upsilon^2)$ and higher have been neglected. The magnitude $d^{g}$ and direction $\mathbf{n}^{g}$ of the vector $\mathbf{p}^g$ is given as
\begin{equation}\label{eq:talreja_crack3}
    d^g=\norm{(J^d_m\,\mathbf{G}_m\,d\mathbf{a}_m)^{g}}\quad\text{and,}\quad \mathbf{n}^g=\dfrac{(\,\mathbf{G}_m\,d\mathbf{a}_m)^{g}}{\norm{(\,\mathbf{G}_m\,d\mathbf{a}_m)^{g}}}
\end{equation}
where the norm of a vector $\mathbf{v}$ is defined as $\norm{\mathbf{v}}=\sqrt{\mathbf{v}\cdot\mathbf{v}}$. Since we assume a continuous distribution of material defects, it is sufficient to consider the vector $\mathbf{p}^g$ as a measure of matrix crack density in our framework. On the other hand, Talreja~(1985)~\cite{talreja1985continuum} considered a statistical distribution of cracks and therefore, a volume-averaged crack density $\mathbf{V}^{(r)}$ was chosen as an internal variable. For a crack plane with orientation $r$ and unit normal vector to the crack plane $\mathbf{n}^{(r)}$, the vector field $\mathbf{V}^{(r)}$ was defined as
\begin{equation}\label{eq:talreja_crack_4}
    \mathbf{V}^{(r)}=D^{(r)}\,\mathbf{n}^{(r)}\quad\text{with}\quad D^{(r)}=\dfrac{1}{\mathcal{V}}\,\sum_{r} d^{(r)}.
\end{equation}
For the sake of completion, the vector field $\mathbf{V}^{(r)}$ can be written in terms of our matrix crack density tensor $\mathbf{G}_m$ using Eq.~\eqref{eq:talreja_crack3} as
\begin{equation}\label{eq:talreja_crack5}
    \mathbf{V}^{(r)}=\left(\dfrac{1}{\mathcal{V}}\,\sum_{r} \norm{(J^d_m\,\mathbf{G}_m\,d\mathbf{a}_m)^{(r)}}\right)\,\dfrac{(\mathbf{G}_m\,d\mathbf{a}_m)^{(r)}}{\norm{(\mathbf{G}_m\,d\mathbf{a}_m)^{(r)}}}.
\end{equation}

For debonding and interfacial slip, a shear lag model such as Hsueh~(1990)~\cite{hsueh1990interfacial} is widely used. According to this model, when the maximum shear stress $\tau$ across a matrix-fiber interface for a given applied load exceeds the shear strength of the interface $\tau_m$, i.e., $\tau > \tau_m$, debonding and interfacial slip initiates. In view of the physical interpretation of Eq.~\eqref{eq:evolution3}, the interfacial shear force in our framework takes the form of the driving force associated with the interfacial slip rate, viz.,  $-\rho_0\,{\partial\overline{\psi}}/{\partial\overline{\boldsymbol{\Lambda}}}$.
Thus, the condition for the initiation of debonding is written as 
\begin{equation}
    \norm{-\rho_0\,\frac{\partial\overline{\psi}}{\partial\overline{\boldsymbol{\Lambda}}}}\ge\tau_m\,\pi df
\end{equation}
where $d$ is the fiber diameter and $\pi\,df$ represents the surface area along which the slippage occurs. A different approach was taken by Talreja~(1991)~\cite{talreja1991continuum} to model the interfacial slip in which a fiber, embedded in a matrix, was pulled out such that the contact between the matrix and the fiber remains intact. For this problem, the interfacial slip is denoted by the vector $\mathbf{d}=b\,\mathbf{s}$ where $b$ and $\mathbf{s}$ represent the magnitude and direction of the interfacial slip vector, respectively. Thereafter, the interfacial slip was characterized by the slippage vector $\boldsymbol{\xi}$ whose only nonzero component along the direction $\mathbf{s}$ is given by
\begin{equation}\label{eq:talreja_debonding1}
    \xi_1=-2\bar{b}\,\pi\,df\quad\text{with}\quad 2\bar{b}=b^++b^-
\end{equation}
$b^{+/-}$ are the magnitudes of the slip vector $\mathbf{d}$ along the two bounding curves of the de-bonded region. Now, to express the component of the slippage vector $\xi_1$ in terms of our interfacial slip vector $\boldsymbol{\Lambda}$, we first observe that the interfacial slip direction is along the tangential direction of the interface. While we consider motions of both the matrix and the fiber phases in~\S~\ref{sec:debonding}, here the fiber was pulled out with a slip vector $\mathbf{d}$ keeping the matrix stationary. In other words, the slip vector $\mathbf{d}$ is directly related to the tangential component of the relative velocity between the two phases across the matrix-fiber interface. Hence, we can write
\begin{equation}\label{eq:talreja_debonding2}
    \mathbf{d}=\int_0^t\boldsymbol{\Lambda}\,dt=\overline{\boldsymbol{\Lambda}}\quad \text{with}\quad b=\norm{\mathbf{d}}.
\end{equation}
Combining Eqs.~\eqref{eq:talreja_debonding1} and~\eqref{eq:talreja_debonding2}, the slippage vector $\boldsymbol{\xi}$ can be expressed in terms of our interfacial slip rate $\boldsymbol{\Lambda}$.

In the case of delamination, we compare our measure of delamination $\hat{\boldsymbol{\Sigma}}$ with the existing models of Rybicki and Kanninen~(1977)~\cite{rybicki1977finite} and Zou~\textit{et al.}~(2001)~\cite{zou2001mode}. Let us consider a crack of length $\Delta a$ in the interface between the two lamin\ae. We choose a Cartesian coordinate system on this interface such that the $x-$ and $y-$ axes are mutually orthogonal base vectors that span the surface of the interface, while the $z-$ axis is perpendicular to the interface plane and along the direction of the separation of the lamin\ae. Based on the virtual crack closure technique, they derived the energy release rates for the three modes of delamination as 
\begin{align}
    G_I&=\lim_{\Delta a\rightarrow0}\int_0^{\Delta a}\frac{1}{2\Delta\,a}\sigma_{zz}(\Delta a-r,0)\,\bar{w}(r)\,dr\sim\frac{1}{2\Delta\,a}Z_{cd}(w_c-w_d), \tag{94a}\label{eq:G1} \\ 
    G_{II}&=\lim_{\Delta a\rightarrow0}\int_0^{\Delta a}\frac{1}{2\Delta\,a}\sigma_{xz}(\Delta a-r,0)\,\bar{u}(r)\,dr\sim\frac{1}{2\Delta\,a}X_{cd}(u_c-u_d), \tag{94b}\label{eq:G2}\\ 
    G_{III}&=\lim_{\Delta a\rightarrow0}\int_0^{\Delta a}\frac{1}{2\Delta\,a}\sigma_
    {yz}(\Delta a-r,0)\,\bar{v}(r)\,dr\sim\lim_{\Delta a\rightarrow0}\frac{1}{2\Delta\,a}Y_{cd}(v_c-v_d).\tag{94c}\label{eq:G3}
\end{align}
\setcounter{equation}{95}
Here $G_I,\,G_{II}$ and $G_{III}$ are the energy release rates for mode I, mode II and III delamination with $\Delta a$ being the crack length. $\sigma_{xz}$, $\sigma_{yz}$ and $\sigma_{zz}$ represent the respective stress components whereas $u$, $v$ and $w$ represent the relative sliding and crack opening respectively at a point on the interface. The last of Eqs.~\eqref{eq:G1}--\eqref{eq:G3} are the finite element approximations of the preceding analytical solutions at nodes $c$ and $d$ located on the interface. $X_{cd}$, $Y_{cd}$ and $Z_{cd}$ are the nodal forces required to close the crack. To compare this model with our damage density due to delamination, let us consider a homogeneous deformation map $x=X+u(L)\,Z$, $y=Y+v(L)\,Z$ and $z=w(L)\,Z$ on any point on the interface where $L$ denotes the distance from the crack tip along the interface. This deformation map yields 
\begin{equation}\label{eq:del_deformation_gradient}
[{F}_{ij}]=\begin{bmatrix}
1 & 0 & u \\
0 & 1 & v \\
0 & 0 & w
\end{bmatrix}
\end{equation}
Without loss of generality, we assume that the inelastic deformation gradient $\mathbf{F}^i$ has the same form as that of $\mathbf{F}$. Now using Eq.~\eqref{eq:nortmal_tangent_part}, the nonzero components of the damage density due to delamination $\hat{\boldsymbol{\Sigma}}$ can be written as 
\begin{equation}\label{eq:components_del}
\hat{{\Sigma}}_{13}=\llbracket{F}^i\rrbracket_{{13}}=u^i_\xi-u^i_\eta,\quad\hat{{\Sigma}}_{23}=\llbracket{F}^i\rrbracket_{{23}}=v^i_\xi-v^i_\eta,\quad\hat{{\Sigma}}_{33}=\llbracket{F}^i\rrbracket_{{33}}=w^i_\xi-w^i_\eta.
\end{equation}
Now following Eshelby's procedure~\cite{eshelby1951force,rajagopal2005role,paul2022use}, the components of the driving/configurational force associated with delamination, $\boldsymbol{\mathcal{D}}$ can be written as
\begin{equation}\label{eq:delamination_driving_force}
    \mathcal{D}_{13}=-\rho_0\frac{\partial\psi}{\partial\llbracket{F}\rrbracket}_{13},\qquad \mathcal{D}_{23}=-\rho_0\frac{\partial\psi}{\partial\llbracket F \rrbracket}_{23},\qquad \mathcal{D}_{33}=-\rho_0\frac{\partial\psi}{\partial\llbracket{F}\rrbracket}_{33}.
\end{equation}
A similar expression for the driving forces can also be found in Mosler and Scheider~(2011)~\cite{mosler2011thermodynamically}. It is evident that the driving force $\boldsymbol{\mathcal{D}}$ has only three nonzero components-- $\mathcal{D}_{13}$, $\mathcal{D}_{23}$ and $\mathcal{D}_{33}$. Moreover, these three components are the forces required to close the crack at a point on the interface. Hence, Eq.~\eqref{eq:delamination_driving_force} acts as a traction-separation law when a suitable form for the Helmholtz free energy $\psi$ is chosen in terms of the jump in the displacements. Now following the notation of Zou~\textit{et al.}~(2001)~\cite{zou2001mode}, we can write $\mathcal{D}_{33}\simeq Z_{cd}$,  $\mathcal{D}_{13}\simeq X_{cd}$ and,   $\mathcal{D}_{23}\simeq Y_{cd}$. Let us assume a linear traction-separation law such that $\psi=\bar{\psi}(\mathbf{C},\mathbf{C}^i,\overline{\mathbf{G}}_{\alpha},\overline{\boldsymbol{\Lambda}})-\psi_{del}(\hat{\boldsymbol{\Sigma}})$ with $\psi_{del}(\hat{\boldsymbol{\Sigma}})=(1/2) C_{ijkl}\, \hat{\Sigma}_{ij}\,\hat{\Sigma}_{kl}$. From the discussion in \S~\ref{sec:constitutive}, the energy dissipated \emph{solely} due to delamination can be written as
\begin{equation}\label{eq:rate_of_dissipation_del}
    \bar{\Pi}_{del}=\frac{1}{2}\,\boldsymbol{\mathcal{D}}\boldsymbol{:}{\hat{\mathbf{\Sigma}}}=-\frac{1}{2}\,\rho_0\dfrac{\partial \psi}{\partial\hat{\mathbf{\Sigma}}}\boldsymbol{:} {\hat{\mathbf{\Sigma}}}.
\end{equation}
With the help of Eqs.~\eqref{eq:components_del}, \eqref{eq:delamination_driving_force} and \eqref{eq:rate_of_dissipation_del}, the energy dissipated due to delamination can be written as a sum of its mode-wise components as
\begin{equation}\label{eq:pi_del1}
\bar{\Pi}_{del}=\underbrace{\frac{1}{2}\mathcal{D}_{33}(w^i_\xi-w^i_\eta)}_{\text{Mode}\, I}\,+\,\underbrace{\frac{1}{2}\mathcal{D}_{13}(u^i_\xi-u^i_\eta)}_{\text{Mode}\, II}\,+\,\underbrace{\frac{1}{2}\mathcal{D}_{23}(v^i_\xi-v^i_\eta)}_{\text{Mode}\, III}.
\end{equation}
Note that $\bar{\Pi}_{del}$ in Eq.~\eqref{eq:pi_del1} represents a \emph{local} energy dissipation at a point on the interface between the lamin\ae~ $\xi$ and $\eta$. Now, following the definition of Rybicki and Kanninen~(1997)~\cite{rybicki1977finite}, the stress intensity factors for different modes of delamination are written as
\begin{equation}\label{eq:global_mode_1,2}
\begin{aligned}
    G_I&=\lim_{\Delta a \to 0}\frac{1}{\Delta\,a}\int^{\Delta a}_0 \bar{\Pi}_I\,dL=\lim_{\Delta a \to 0}\frac{1}{2\Delta\,a}\int^{\Delta a}_0\mathcal{D}_{33}(L)\,[w^i_\xi(L)-w^i_\eta(L)]\,dL,\\ 
    G_{II}&=\lim_{\Delta a \to 0}\frac{1}{\Delta\,a}\int^{\Delta a}_0 \bar{\Pi}_{II}\,dL=\lim_{\Delta a \to 0}\frac{1}{2\Delta\,a}\int^{\Delta a}_0\mathcal{D}_{13}(L)\,[u^i_\xi(L)-u^i_\eta(L)]\,dL,\\ 
    G_{III}&=\lim_{\Delta a \to 0}\frac{1}{\Delta\,a}\int^{\Delta a}_0 \bar{\Pi}_{III}\,dL=\lim_{\Delta a \to 0}\frac{1}{2\Delta\,a}\int^{\Delta a}_0\mathcal{D}_{23}(L)\,[v^i_\xi(L)-v^i_\eta(L)]\,dL.
\end{aligned}
\end{equation}

\subsection{Possible methods for experimental quantification of the derived damage measures}

From the expressions of the developed damage measures, it can be observed that these kinematic variables can be characterized through measurements of displacement fields in appropriate experimental setups. In this section, we briefly discuss possible experimental methods that can be used for quantification of the derived damage measures from suitable experimental data.

Let us first consider a special case in which matrix cracking is the predominant mode of failure and the effects of other damage mechanisms are ignored. Under this condition, the microstructural changes in the body is governed by matrix cracking only and hence, one can write $\mathbf{F}^i\simeq \mathbf{F}^d_m$ and $\mathbf{F}^d_f\simeq \mathbf{I}$. Furthermore, in view of Eq.~\eqref{eq:constitutent_def_grad}, the other tangent maps can be written as $\mathbf{F}^r_f\simeq\mathbf{F}^d_m$ and $\mathbf{F}^r_m\simeq\mathbf{I}$. Physically, these conditions imply that the configurations $\kappa^d_m(\mathcal{B}_0)$ and $\kappa^d_f(\mathcal{B}_0)$  coincide with the natural configuration $\kappa_i(\mathcal{B}_0)$ and the reference configuration $\kappa_R(\mathcal{B}_0)$, respectively. Although the derived crack density tensor $\mathbf{G}_m$ cannot be \emph{directly} measured from the existing experimental setups, it can be determined through post-experimental analysis of full-field displacement measurements. Under the prescribed conditions, the crack density tensor $\mathbf{G}_m$ can be obtained from the `Curl' of the tensor field $\mathbf{F}^i$ (or alternatively, $\mathbf{F}$ and $\mathbf{F}^e$), which in turn can be determined from the gradients of the experimentally measured displacement fields. An \emph{in situ} tensile test using X-ray computed microtomography (micro-CT) with digital image correlation (DIC) or digital volume correlation (DVC) techniques~\cite{schilling2005x,lee2022situ} is particularly suitable for this purpose. In a similar manner, the damage measure for fiber breakage, $\mathbf{G}_f$ can be obtained from the displacement fields in a fragmentation test that can be measured using an \emph{in situ}~\cite{garcea2017mapping} or \emph{ex situ}~\cite{drouhet20233d} micro-CT analysis, coupled with DIC and DVC techniques.

To obtain the damage measure, $\boldsymbol{\lambda}_m$ corresponding to debonding and interfacial slip, it is sufficient to compute the tangential velocity along the matrix-fiber interface. This can be achieved by performing a fiber pull-out test coupled with DIC technique that enables the measurement of the required displacement fields at different time steps~\cite{tabiai2018situ}. More accurate results could also be obtained through a micro-CT analysis coupled with a DVC technique~\cite{janicke2022debonding} during a fiber pull-out test. To characterize delamination, typically a double cantilever beam (DCB), a mixed-mode bending (MMB), or an end-notched flexure (ENF) test is performed. Since a full displacement field is required for the damage measure $\hat{\mathbf{\Sigma}}$, a DCB test coupled with a DIC technique~\cite{zhu2020digital} is deemed to be useful for this purpose. For better accuracy, a micro-CT analysis~\cite{tsokanas20253d} can also be performed. It is important to note here that in our discussion of possible experimental methods, we considered that different damage mechanisms can be decoupled during different tests. Although certain experimental conditions may satisfy this requirement, it may not always be possible to avoid interactions between different damage mechanisms completely. To avoid this issue, new experimental techniques need to be proposed. Moreover, exact experimental protocols need to be developed for the quantification of the derived damage measures. Both these tasks are out of the scope of the current paper and need to be addressed in the future. 
}

\section{Concluding remarks}\label{Conclusion}
The paper presents a novel kinematic framework for characterizing damage content in fiber-reinforced composite materials undergoing large deformation using the central idea of both multiple natural configurations and multi-continuum theory. In the context of multiple natural configurations, we employ $\mathbf{F}=\mathbf{F}^e\,\mathbf{F}^i$, whereas in the spirit of multi-continuum theory, this $\mathbf{F}^i$ is further decomposed into as $\mathbf{F}^i=\mathbf{F}^r_\alpha\,\mathbf{F}^d_\alpha$ where $\alpha$ represent two constituent phases. The framework is suitable for any multiphase material. Using this kinematic framework, we characterized the damage associated with four mechanisms observed in fiber-reinforced laminated composites. These are matrix cracking, fiber breakage, interfacial slip and debonding and delamination. The measure of damage are obtained by using physical arguments such as closure failure of a circuit, relative tangential velocity along the interface, jump in the displacement field as well as their geometric interpretations in terms of local torsions in the material manifolds. The final forms of these damaged contents are listed below:\\
For matrix cracks,
\begin{equation}
    \mathbf{G}_m=\dfrac{1}{J^d_m}\,\mathbf{F}^{d}_m\,(\text{Curl}\,\mathbf{F}^d_m)=J^e\,J^r_m\,\mathbf{F}^{e^{-1}}\,\mathbf{F}^{r^{-1}}_m\,\text{curl}\,(\mathbf{F}^{e^{-1}}\,\mathbf{F}^{r^{-1}}_m).
\end{equation}
For fiber breakage,
\begin{equation}
    \mathbf{G}_f=\dfrac{1}{J^d_f}\,\mathbf{F}^{d}_f\,(\text{Curl}\,\mathbf{F}^d_f)=J^e\,J^r_f\,\mathbf{F}^{e^{-1}}\,\mathbf{F}^{r^{-1}}_f\,\text{curl}\,(\mathbf{F}^{e^{-1}}\,\mathbf{F}^{r^{-1}}_f).
\end{equation}
For debonding and interfacial slip,
\begin{equation}
    \boldsymbol{\lambda}_m=\mathbf{F}^d_m\,\mathbf{V}_{rel}=\mathbf{v}^d_m\,-\mathbf{F}^d_m\mathbf{F}^{d^{-1}}_f\mathbf{v}^d_f.
\end{equation}
And finally for delamination,
\begin{equation}
{\hat{\mathbf{\Sigma}}=\llbracket\mathbf{F}^i\rrbracket,}\quad \tilde{\mathbf{\Sigma}}_\delta=\mathbf{\Sigma}^T\boldsymbol{\vartheta}\,\tilde{\mathbf{F}}^{i^{-1}}_\delta=\boldsymbol{\sigma}^T\boldsymbol{\vartheta}\,\tilde{\mathbf{F}}^{e}_\delta,\quad\delta=\xi,\eta.   
\end{equation}
This work can be further extended to study the {specific} constitutive behavior, initiation and evolution of damage {in different types of laminated composites using pertinent numerical and experimental techniques}. 

\section*{Acknowledgments}
The authors gratefully acknowledge the financial support from Anusandhan National Research Foundation (ANRF), Government of India for this research through the grant \# SRG/2023/000197.

\bibliographystyle{acm}

\setcounter{section}{0}
\renewcommand{\thesection}{Appendix\,\Alph{section}}
\setcounter{equation}{0}
\renewcommand{\theequation}{A.\arabic{equation}}

{
\section{Derivation of the relation between the crack density tensor and Curl\,($\mathbf{F}$)}

\subsection{Derivation of the jump in the displacement field}\label{discontinious field}\label{sec:Appendix_discontinious_field}

Valanis and Panoskaltsis~(2005)~\cite{valanis2005material} showed that the jump in the displacement field across a crack surface can be written as $\text{Curl}(\mathbf{F})$ which we use in our derivation of the crack density tensor for matrix cracking and fiber breakage in \S~\ref{sec:matrix}. For the sake of completion, this derivation is provided here. 

Let us first consider a two dimensional deformation such that $\boldsymbol{\Delta}X^1\rightarrow\boldsymbol{\Delta}x^1$, $\boldsymbol{\Delta}X^2\rightarrow\boldsymbol{\Delta}x^2$. In an undeformed body, one can reach point $P$ starting from the point $R$ through two paths: the path $RSP$ that requires a vector sum of two displacements $(\boldsymbol{\Delta}X^1,0)$ and $(0,\boldsymbol{\Delta}X^2)$ and path $RQP$ which requires a displacement $(0,\boldsymbol{\Delta}X^2)+(\boldsymbol{\Delta}X^1,0)$. Now in the deformed configuration of the body, the path corresponding to $RQ$, denoted by $\boldsymbol{\Delta}x^l_{(1)}$, can be written as
\begin{equation}\label{eq:local_map_1}
\boldsymbol{\Delta}x^l_{(1)}={F}^l_1(0,0)\boldsymbol{\Delta}X^1.
\end{equation}
Since the path $QP$ originates from the point $Q$, the displacement vector along the path $QP$ can be written as 
\begin{equation} \label{eq:local_map_2}
\boldsymbol{\Delta}x^l_{(2)}={F}^l_2(\boldsymbol{\Delta}X^1,0)\boldsymbol{\Delta}X^2={F}^l_2(0,0)\boldsymbol{\Delta}X^2+{F}^l_{2,1}(0,0)\boldsymbol{\Delta}X^1 \boldsymbol{\Delta}X^2.
\end{equation}
In a similar manner, the position vector of the point $P$ following the path $RSP$ can be written as
\begin{equation} \label{eq:local_map_3}
\boldsymbol{\Delta}x^l_{(3)}={F}^l_2(0,0)\boldsymbol{\Delta}X^2
\end{equation}
and,
\begin{equation}\label{eq:local_map_4}
\boldsymbol{\Delta}x^l_{(4)}={F}^l_1(0,0)\boldsymbol{\Delta}X^1+{F}^l_{1,2}(0,0)\boldsymbol{\Delta}X^1 \boldsymbol{\Delta}X^2
\end{equation}
where $\boldsymbol{\Delta}x^l_{(3)}$ and $\boldsymbol{\Delta}x^l_{(4)}$ are the displacement vectors along the paths $RS$ and $SP$ respectively. Now, in the absence of a crack at point $P$, its position vectors obtained via two different paths have to be the same and hence, it must satisfy the compatibility condition
\begin{equation}\label{eq:continiuity_body}
    \boldsymbol{\Delta}x^l_{1}+ \boldsymbol{\Delta}x^l_{2}=  \boldsymbol{\Delta}x^l_{3}+\boldsymbol{\Delta}x^l_{4}.
\end{equation}
However, when a crack exists at point $P$, it can be characterized by the jump in the displacement field at that point as shown in Fig.~\ref{fig:Crack}. The jump in the displacement field can be written as
\begin{equation}
    p^l=\boldsymbol{\Delta}x^l_{3}+\boldsymbol{\Delta}x^l_{4}-\boldsymbol{\Delta}x^l_{1}+ \boldsymbol{\Delta}x^l_{2}=({F}^l_{1,2}(0,0)-{F}^l_{2,1}(0,0))\boldsymbol{\Delta}X^1\,\boldsymbol{\Delta}X^2
\end{equation}
Now, carrying out the same exercise for a 3-D deformation, one can easily find the jump in the displacement field as
\begin{equation}
p^l= ({F}^l_{1,2}-{F}^l_{2,1})\boldsymbol{\Delta}X^1\,\boldsymbol{\Delta}X^2+({F}^l_{2,3}-{F}^l_{3,2})\boldsymbol{\Delta}X^2\,\boldsymbol{\Delta}X^3+({F}^l_{3,1}-{F}^l_{1,3})\boldsymbol{\Delta}X^3\,\boldsymbol{\Delta}X^1=\epsilon^q_{jp}\,F^l_{q,p}\,\boldsymbol{\Delta}A^j
\end{equation}
where $\epsilon^j_{pq}$ is the Levi-Civita permutation tensor and $\boldsymbol{\Delta}A$ is the area spanned by the displacement vectors $\Delta X^i$ such that $\Delta A^j=\epsilon^j_{mn}\,\Delta X^m\Delta X^n$. Therefore, one can write the jump in the displacement vector as
\begin{equation}
    \mathbf{p}=(\text{Curl}\,\mathbf{F})\, d\mathbf{A}.
\end{equation}

\subsection{Relation with matrix crack density tensor $\mathbf{G}_m$ with the geometric interpretation}\label{sec:Appendix_matrix_crack}

To establish the relation between the matrix crack density tensor $\mathbf{G}_m$ in Eq.~\eqref{eq:matrix_crack_density} and the corresponding torsion $\mathbf{T}_m$ in Eq.~\eqref{eq:torsion_m}, we start with the definition of the crack density tensor $\mathbf{G}_m$. From Eq.~\eqref{eq:matrix_crack_density}, one can write
\begin{equation}\label{eq:defination_curl_appendix}
J^d_m\,\mathbf{F}^{d^{-1}}_m\,\mathbf{G}_m\big|^i_{j}=\text{Curl}\,\mathbf{F}^d_m\big|^i_j=\epsilon^s_{jr}\frac{\partial{F}^{d\,^i}_{m\,_{s}}}{\partial X^r}=\epsilon^s_{jr}\partial_r {F}^{d\,^{\,i}}_{m{\,s}}.
\end{equation}
Now let us fix the index $j$ in the expression of $J^d_m\,\mathbf{F}^{d^{-1}}_m\,\mathbf{G}_m$, in Eq.~\eqref{eq:defination_curl_appendix}. Using the properties of Levi-Civita permutation tensor, we get
\begin{equation}\label{eq:fixed_index}
J^d_m\,\mathbf{F}^{d^{-1}}_m\,\mathbf{G}_m\big|^i_{1}=\text{Curl}\,(\mathbf{F}^d_m)\big|^i_{1}=\epsilon^3_{12}\,\partial_2{F}^{d^{\,i}}_{m\,_{3}}+\epsilon^2_{13}\,\partial_3{F}^{d^{\,i}}_{m_{\,2}}=\partial_2{F}^{d^{\,i}}_{m_{\,3}}-\partial_3{F}^{d^{\,i}}_{m_{\,2}}.
\end{equation}
One can easily notice that the right hand side of Eq.~\eqref{eq:fixed_index} is the same as the component $2\mathbf{F}^{d}_{m}\mathbf{T}_m\big|{^i_{23}}$ in Eq.~\eqref{eq:torsion_m}. Similarly, other components of the torsion tensor $\mathbf{T}_m$ can be shown to be related to the components $\text{Curl}(\mathbf{F}^d_m)$. Hence, there exists a one-to-one correspondence between the components of the torsion tensors $\mathbf{T}_{\alpha}$ and the crack density tensors $\mathbf{G}_{\alpha}$.}

\section{Derivation of total damage content due to delamination and true measure of the damage density}\label{sec:appendix_del_total}

In this section, we show the detailed derivation of the damage content due to delamination and the true measure of the damage density. We start with the damage content density in Eq.~\eqref{eq:tangential_mismatch} as
\begin{equation}\label{eq:tangential_mismatch2}
    \mathbf{\Sigma}^T\tilde{\mathbf{t}}_1=\llbracket{\mathbf{F}^i}\rrbracket\,(\tilde{\mathbf{t}}_1\times\tilde{\mathbf{N}}).
\end{equation}
\begin{equation}\label{eq:tangential_mismatch4}
   \llbracket{\mathbf{F}^i}\rrbracket\,(\tilde{\mathbf{t}}_1\times(\tilde{\mathbf{t}}_1\times\tilde{\mathbf{t}}_2))=-\llbracket\mathbf{F}^i\rrbracket\tilde{\mathbf{t}}_2.
\end{equation}
Since the infinitesimal fiber $d\mathbf{X}$ can be written as $d\mathbf{X}=\tilde{\mathbf{t}}_2\,d{L}$, the total damage content due to delamination can be evaluated using the Eqs.~\eqref{eq:burgers_interface_reference} and \eqref{eq:tangential_mismatch4} as
\begin{equation}
    \mathbf{b}^R= \int_{\Omega}(\text{Curl}\,\mathbf{F}^i)^T\,\mathbf{N}\,dA-\int_\Gamma\llbracket\mathbf{F}^i\rrbracket\,\tilde{\mathbf{t}}_2\,d\mathbf{L}=\int_{\Omega}(\text{Curl}\,\mathbf{F}^i)^T\,\mathbf{N}\,dA+\int_\Gamma  \mathbf{\Sigma}^T\tilde{\mathbf{t}}_1\,d{L}.
\end{equation}
Now, using the identity operator $\mathbf{1}=\mathbf{N}\otimes\mathbf{N}+\tilde{\mathbf{t}}_1\otimes\tilde{\mathbf{t}}_1+\tilde{\mathbf{t}}_2\otimes\tilde{\mathbf{t}}_2$, and taking the projection of $\llbracket\mathbf{F}^i\rrbracket$ onto itself, we obtain
\begin{equation}
    \llbracket\mathbf{F}^i\rrbracket=\llbracket\mathbf{F}^i\rrbracket\mathbf{1}
\end{equation}
\begin{equation}\label{eq:projection}
\implies\llbracket\mathbf{F}^i\rrbracket=\llbracket\mathbf{F}^i\rrbracket(\tilde{\mathbf{N}}\otimes\tilde{\mathbf{N}})+\llbracket\mathbf{F}^i\rrbracket(\tilde{\mathbf{t}}_1\otimes\tilde{\mathbf{t}}_1)+\llbracket\mathbf{F}^i\rrbracket(\tilde{\mathbf{t}}_2\otimes\tilde{\mathbf{t}}_2).
\end{equation}
Thus, Eq.~\eqref{eq:tangential_mismatch4} can be also rewritten as 
\begin{equation}\label{eq:tangential_mismatch3}
    \mathbf{\Sigma}^T\tilde{\mathbf{t}}_2=\llbracket{\mathbf{F}^i}\rrbracket\,(\tilde{\mathbf{t}}_2\times\tilde{\mathbf{N}}).
\end{equation}
Now using Eqs.~\eqref{eq:tangential_mismatch3} and \eqref{eq:tangential_mismatch4}, we can rewrite the Eq.~\eqref{eq:projection} as
\begin{equation}    
\llbracket\mathbf{F}^i\rrbracket=\mathbf{a}\otimes\tilde{\mathbf{N}}-\mathbf{\Sigma}^T(\tilde{\mathbf{t}}_1\otimes\tilde{\mathbf{t}}_2-\tilde{\mathbf{t}}_2\otimes\tilde{\mathbf{t}}_1)
\end{equation}
\begin{equation}\label{eq:nortmal_tangent_part2}
   \implies\llbracket{\mathbf{F}^i}\rrbracket\,=\mathbf{F}^i_\xi\,-\mathbf{F}^i_\eta= \mathbf{a}\otimes\tilde{\mathbf{N}}-\mathbf{\Sigma}^T\boldsymbol{\vartheta}.
\end{equation}
where $\mathbf{a}$ is an arbitrary vector and $\boldsymbol{\vartheta}=\tilde{\mathbf{t}}_1\otimes\tilde{\mathbf{t}}_2-\tilde{\mathbf{t}}_2\otimes\tilde{\mathbf{t}}_1$. The first of the right-hand side of Eq.~\eqref{eq:nortmal_tangent_part2} accounts for the normal part of $\llbracket\mathbf{F}^i\rrbracket$, whereas the second accounts for its tangential part. Now, to obtain the projection of the jump on the two-dimensional interface, we define the projection tensor $\tilde{\mathbf{1}}$ such that
\begin{equation}    (\mathbf{F}^i)\tilde{\mathbf{1}}=\tilde{\mathbf{F}}^i,\quad\text{where}\quad\tilde{\mathbf{1}}=\tilde{\mathbf{t}}_1\otimes\tilde{\mathbf{t}}_1+\tilde{\mathbf{t}}_2\otimes\tilde{\mathbf{t}}_2.
\end{equation}
Now projecting $\llbracket\mathbf{F}^i\rrbracket$ in Eq.~\eqref{eq:nortmal_tangent_part2} onto $\tilde{\mathbf{1}}$, we get
\begin{equation} 
-\mathbf{\Sigma}^T \boldsymbol{\vartheta}=\llbracket{\mathbf{F}^i}\rrbracket\,\tilde{\mathbf{1}}=\tilde{\mathbf{F}}^i_\xi\,-\tilde{\mathbf{F}}^i_\eta.
\end{equation}
Here $\tilde{\mathbf{F}}$ is the tangent map projected on the interface. It is important to note that the interface damage density in Eqs.~\eqref{eq:burgers_interface_reference} and Eq.~\eqref{eq:burgers_interface_current} depends on the choice of configuration. To provide a measure of damage density as invariant under compatible changes in the reference configuration, we consider a smooth tangent map $\mathbf{H}$ between the two reference configurations such that $d\mathbf{X}_2=\mathbf{H}\,d\mathbf{X}_1$. Since under compatible deformation, the incompatibility for both the reference configurations have to be the same, we can write
\begin{equation}\label{eq:curl_equal_ref}
  \int_{\Omega_1}(\text{Curl}_1\,\mathbf{F}^i_1)^T\,\mathbf{N}_{1}\,dA_1-\int_{\Gamma_1}\llbracket\mathbf{F}^i_1\rrbracket d\mathbf{X}_1=\int_{\Omega_2}(\text{Curl}_2\,\mathbf{F}^i_2)^T\,\mathbf{N}_{2}\,dA_2-\int_{\Gamma_2}\llbracket\mathbf{F}^i_2\rrbracket d\mathbf{X}_2.
\end{equation}
Since bulk interface damage density far from the interface is invariant, the area integral in the first two terms on the left and right hand sides of Eq.~\eqref{eq:curl_equal_ref} are equal. Therefore, Eq.~\eqref{eq:curl_equal_ref} reduces to
\begin{equation}\label{eq:equal_ref1}
  \int_{\Gamma_1}\llbracket\mathbf{F}^i_1\rrbracket d\mathbf{X}_1=\int_{\Gamma_2}\llbracket\mathbf{F}^i_2\rrbracket d\mathbf{X}_2.
\end{equation}
Let $\tilde{\mathbf{t}}_{\boldsymbol{\beta}}$ and ${L}_{\boldsymbol{\beta}}$ be the unit tangent vector and the length of the curve $\Gamma_{\boldsymbol{\beta}}$ respectively where $\boldsymbol{\beta}= 1\,,2$ such that $\tilde{\mathbf{t}}_2\,d{L}_2=d\mathbf{X}_2$. As we discussed in the \S~\ref{sec:Continuous and discontinuous manifold} the circuit is independent of path integral. Hence, Eq.~\eqref{eq:equal_ref1} can be written as 
\begin{equation}\label{eq:tangential_compatibilty}
    \llbracket\mathbf{F}^i_2\rrbracket\mathbf{H}\,\tilde{\mathbf{t}}_1- \llbracket\mathbf{F}^i_1\rrbracket\,\tilde{\mathbf{t}}_1=0.
\end{equation}
Eq.~\eqref{eq:tangential_compatibilty} must be satisfied to maintain the compatibility between two configurations. Moreover, since $\tilde{\mathbf{t}}_1$ is arbitrary on the tangent plane of the surface $S_R$, we obtain. 
\begin{equation}\label{eq:tangential_compatibilty_2}
   (\llbracket\mathbf{F}^i_2\rrbracket\mathbf{H}- \llbracket\mathbf{F}^i_1\rrbracket) \tilde{\mathbf{t}}_1=0.
\end{equation}
Now projecting Eq.~\eqref{eq:tangential_compatibilty_2} onto $\tilde{\mathbf{1}}$ and using the compatible deformation $\tilde{\mathbf{H}}=\tilde{\mathbf{F}}_{2_\delta}^{i^{-1}}\,\tilde{\mathbf{F}}_{1_\delta}^{i}$ we can write the Eq.~\eqref{eq:tangential_compatibilty_2} as 
\begin{equation}
    \boldsymbol{\Sigma}_2^T\,\boldsymbol{\vartheta}\,\tilde{\mathbf{F}}_{2_\delta}^{i^{-1}}= \boldsymbol{\Sigma}_1^T\,\boldsymbol{\vartheta}\,\tilde{\mathbf{F}}_{1_\delta}^{i^{-1}}.
\end{equation}
The equality shows the invariance of the obtained damage density. Finally, we can write the true damage density as
\begin{equation}\label{eq:true_interface_dislocation2}
    \tilde{\mathbf{\Sigma}}_\delta=\mathbf{\Sigma}^T\boldsymbol{\vartheta}(\tilde{\mathbf{F}}^{i^{-1}})_\delta,\quad\delta=\xi,\eta.
\end{equation}

\section{Derivations used in the geometric interpretation of delamination in~\S~\ref{sec:Geometric interpretation of the measures of damage}}

\subsection{Derivation of $\partial\mathbf{F}^{i}$}\label{sec:appendixC}
Here we provide a proof of Eq.~\eqref{eq:convected_derivative_F_i}. A differentiation of Eq.~\eqref{eq:jump_F_i} with respect to the referential coordinate can be written as 
\begin{equation}\label{eq:derivative_jump_F_i}
    \partial_B{F}^{i^{\,a}}_{\,A}= \partial_B\left(h_\xi{F}^{i^{\,a}}_{\xi_{\,A}}\right)+ \partial_B\left(h_\eta{F}^{i^{\,a}}_{\eta_{\,A}}\right).
\end{equation}
Now using product rule, the Eq.~\eqref{eq:derivative_jump_F_i} can be written as 
\begin{equation}\label{eq:convected_derivative_F_i_2}
    \partial_B{F}^{i^{\,a}}_{\,A}=(\partial_B h_\xi){F}^{i^{\,a}}_{\xi_{\,A}}+h_\xi\,\partial_B{F}^{i^{\,a}}_{\xi_{\,A}}+(\partial_B h_\eta){F}^{i^{\,a}}_{\eta_{\,A}}+h_\eta\,\partial_B{F}^{i^{\,a}}_{\eta_{\,A}}.
\end{equation}
where the derivatives of the Heaviside step functions on both sides of the interface $S_R$ follows the identity
\begin{equation}\label{eq:lamda_prop_2}
    \partial_Bh_\xi=n_B\Delta_{S_R},\quad\quad\partial_Bh_\eta=-n_B\Delta_{S_R},\quad\text{and}\quad\partial_B h_\xi=-\partial_B h_\eta.
\end{equation}
Here $\Delta_{S_R}$ is the Dirac delta distribution at the interface $S_R$ and $\mathbf{n}$ is unit normal to the interface $S_R$. Now substituting the Eq.~\eqref{eq:lamda_prop_2} into ~\eqref{eq:convected_derivative_F_i_2} we get
\begin{equation}\label{eq:convected_derivative_F_i_3}     \partial_B{F}^{i^{\,a}}_{\,A}=h_\xi\,\partial_B{F}^{i^{\,a}}_{\xi_{\,A}}+h_\eta\,\partial_B{F}^{i^{\,a}}_{\eta_{\,A}}+n_B\Delta_{S_R}{F}^{i^{\,a}}_{\xi_{\,A}}-n_B\Delta_{S_R}{F}^{i^{\,a}}_{\eta_{\,A}}.
\end{equation}
From the definition of the jump condition Eq.\eqref{eq:jump}, we can write
\begin{equation}\label{eq:def_jump_F_i}
   n_B\Delta_{S_R}{F}^{i^{\,a}}_{\xi_{\,A}}-n_B\Delta_{S_R}{F}^{i^{\,a}}_{\eta_{\,A}}=n_B\Delta_{S_R}\,\llbracket{F}^i\rrbracket^{\,a}_{\,A}.
\end{equation}
Now substituting the Eq.~\eqref{eq:def_jump_F_i} into \eqref{eq:convected_derivative_F_i_3} we obtain
\begin{equation}    \partial_B{F}^{i^{\,a}}_{\,A}=h_\xi\partial_B{F}^{i^{\,a}}_{\xi_{\,A}}+h_\eta\partial_B{F}^{i^{\,a}}_{\eta_{\,A}}+n_B\Delta_{S_R}\,\llbracket{F}^i\rrbracket^{\,a}_{\,A}.
\end{equation}

\subsection{Derivation of the expression for $\mathbf{F}^{i^{-1}}$}\label{sec:appendixB}
In general, the expression for $\mathbf{F}^{i^{-1}}$ used in the Eq.~\eqref{eq:metric_C^i} does not hold. However in this section, we shows that this counter-intuitive expression from $\mathbf{F}^{i^{-1}}$ is indeed correct. Let us define a second order tensor $\mathbf{K}$ such that
\begin{equation}\label{eq:jump_k}
    \mathbf{K}=h_\xi\mathbf{F}^{i^{-1}}_\xi+ h_\eta\mathbf{F}^{i^{-1}}_\eta
\end{equation}
We now evaluate the product of this tensor with $\mathbf{F}^i$. Using the expression of $\mathbf{F}^i$ and $\mathbf{K}$ from the Eq.~\eqref{eq:jump_F_i} and \eqref{eq:jump_k}, we can write
\begin{equation}
\begin{aligned}
    \mathbf{F}^i\,\mathbf{K}&=(h_\xi\mathbf{F}^i_\xi+ h_\eta\mathbf{F}^i_\eta)( h_\xi\mathbf{F}^{i^{-1}}_\xi+ h_\eta\mathbf{F}^{i^{-1}}_\eta)\\&=h_\xi h_\xi\mathbf{F}^i_\xi\mathbf{F}^{i^{-1}}_\xi+ h_\xi h_\eta\mathbf{F}^i_\xi\mathbf{F}^{i^{-1}}_\eta+h_\eta h_\xi\mathbf{F}^i_\eta\mathbf{F}^{i^{-1}}_\xi+h_\eta h_\eta\mathbf{F}^i_\eta\mathbf{F}^{i^{-1}}_\eta.
\end{aligned}
\end{equation}
Now using the properties of $h_\xi$ and $h_\eta$ from Eq,~\eqref{eq:Lambda_prop}, we can write
\begin{equation}\label{eq:heavyside_F_i}
	 \mathbf{F}^i\,\mathbf{K}=h_\xi h_\xi\, \mathbf{F}^{i}_\xi\,\mathbf{F}^{i^{-1}}_\xi+ h_\eta h_\eta\, \mathbf{F}^{i}_\eta\,\mathbf{F}^{i^{-1}}_\eta=\mathbf{I}.
\end{equation}
Therefore, $\mathbf{K}$ is the required inverse of $\mathbf{F}^i$.

{
\subsection{Relation between the damage density due to delamination and its geometric interpretation}\label{sec:Appendix_delamination}

As discussed in \S~\ref{sec:delamination}, the damage density due to delamination across the interface between two lamin\ae~ is written as (Eq.~\eqref{eq:tangential_mismatch}) 
\begin{equation}\label{eq:damage_Appendix}
    \boldsymbol{\Sigma}^T\,\tilde{\mathbf{t}}_1=\llbracket \mathbf{F}^i\rrbracket\,(\tilde{\mathbf{t}}_1\times\tilde{\mathbf{{N}}}).
\end{equation}
In \S~\ref{sec:delamination}, we measured the damage density in a local coordinate system at the interface whose base vectors are given as $\tilde{\mathbf{t}}_1$, $\tilde{\mathbf{t}}_2$ and $\tilde{\mathbf{N}}=\tilde{\mathbf{t}}_1\times\tilde{\mathbf{t}}_2$. To show that the measures of damage density obtained from the physical arguments and the geometric perspective are equivalent, we choose a set of global, Cartesian base vectors $\mathbf{E}_i$. In these base vectors, the representation of Eq.~\eqref{eq:damage_Appendix} takes the form
\begin{equation}\label{eq:damage_interface_appendix_2}
{\Sigma}^T\tilde{{t}}_1\Bigg|_a=\llbracket{F^i}\rrbracket^a_j\,(\tilde{{t}}_1\times\tilde{{N}})^j=\epsilon^j_{pq}\,\llbracket{F^i}\rrbracket^a_j\,\tilde{{t}}\,^p_{1}\tilde{{N}}^q.
\end{equation}
Now expanding Eq.~\eqref{eq:damage_interface_appendix_2} and using the properties of Levi-Civita permutation tensor, we obtain
\begin{equation}\label{eq:damage_interface_appendix_4}    
({\Sigma}^T\tilde{{t}}_1)^a=\tilde{{t}}\,^2_{1}\left( \llbracket{F^i}\rrbracket^a_1\,\tilde{{N}}^3-\llbracket{F^i}\rrbracket^a_3\,\tilde{{N}}^1\right).
\end{equation}
One can observe the similarities between Eq.~\eqref{eq:damage_interface_appendix_4} and the expression of torsion in Eq.~\eqref{eq:tosion_int_del}. Specifically, the component of torsion $T^D_{13}$ across the interface between the two lamin\ae~can be written as
\begin{equation}\label{eq:tosion_int_del_Appendix}
   T^D_{13}\Bigg|_{S_R}=\frac{1}{2}({F}^{i^{-1}})^D_a\,\Delta_{S_R} \tilde{{t}}\,^2_{1}\Bigg(\llbracket{F^i}\rrbracket^a_{1}\,\tilde{{N}}^3-\llbracket{F^i}\rrbracket^a_{3}\,\tilde{{N}}^1\Bigg)=\frac{1}{2}({F}^{i^{-1}})^D_a\,\Delta_{S_R} ({\Sigma}^T\tilde{{t}}_1)^a.
\end{equation}
In a similar manner, other components of the torsion tensor such as $T^D_{23}$ can be shown to be related with $\boldsymbol{\Sigma}^T\,\tilde{\mathbf{t}}_2$. In view of Eq.~\eqref{eq:jump_interface_projection} and the definition of $\boldsymbol{\vartheta}$, since the vectors $\boldsymbol{\Sigma}^T\,\tilde{\mathbf{t}}_1$ and $\boldsymbol{\Sigma}^T\,\tilde{\mathbf{t}}_2$ are the only necessary vectors to define the damage density due to delamination $\tilde{\boldsymbol{\Sigma}}$, it is hence shown that there exists a one-to-one correspondence between the components of the damage density due to delamination $\tilde{\boldsymbol{\Sigma}}$ and the corresponding torsion.
}

\end{document}